\documentclass[journal,10 pt,twocolumn]{IEEEtranTCOM}
\usepackage{epsfig}
\usepackage{graphicx}
\usepackage{caption}
\usepackage{subcaption}
\usepackage{balance}
\usepackage{float}
\usepackage{psfrag}
\usepackage{xcolor}
\usepackage[utf8]{inputenc}
\usepackage{amsfonts,amsmath,amssymb,mathrsfs,txfonts}
\newtheorem{theorem}{Theorem}
\newtheorem{lemma}{Lemma}
\newtheorem{cor}{Corollary}
\newtheorem{definition}{Definition}
\newtheorem{pro}{Proposition}
\newtheorem{remark}{Remark}
\usepackage{url}
\usepackage{cite}
\usepackage{verbatim}
\usepackage{multirow}
\usepackage{multicol}
\usepackage[font=small]{caption}
\usepackage{pifont}
\usepackage{numprint}
\npthousandsep{,}
\npthousandthpartsep{}
\npdecimalsign{.}
\usepackage{lettrine}
\begin{document}
\title{A Rank-One Optimization Framework and its Applications to Transmit Beamforming}
\author{Tuan Anh Le, {\it Senior Member IEEE}, Derrick Wing Kwan Ng, {\it Fellow IEEE}, and Xin-She Yang
\thanks{T. A. Le and X.-S. Yang are with the Faculty of Science and Technology, Middlesex University, London, NW4 4BT, UK. D. W. K. Ng is with the University of New South Wales, Sydney, NSW, Australia. Email: t.le@mdx.ac.uk; w.k.ng@unsw.edu.au; x.yang@mdx.ac.uk. }\thanks{ D. W. K. Ng is supported by the Australian Research Council's Discovery Projects (DP210102169, DP230100603).}
}
\markboth{IEEE Transactions on Vehicular Technology, DOI: 10.1109/TVT.2023.3303623 }%
{Shell \MakeLowercase{\textit{et al.}}: Bare Demo of IEEEtran.cls for IEEE Journals}
\maketitle

\begin{abstract}
This paper proposes an elegant optimization framework consisting of a mix of linear-matrix-inequality and second-order-cone constraints. The proposed framework generalizes the semidefinite relaxation (SDR) enabled solution to the typical transmit beamforming problems presented in the form of quadratically constrained quadratic programs (QCQPs) in the literature. It is proved that the optimization problems subsumed under the framework always admit a rank-one optimal solution when they are feasible and their optimal solutions are not trivial. This finding indicates that the relaxation is tight as the optimal solution of the original beamforming QCQP can be straightforwardly obtained from that of the SDR counterpart without any loss of optimality. Four representative examples of transmit beamforming, i.e., transmit beamforming with perfect channel state information (CSI), transmit beamforming with imperfect CSI, chance-constraint approach for imperfect CSI, and reconfigurable-intelligent-surface (RIS) aided beamforming, are shown to demonstrate how the proposed optimization framework can be realized in deriving the SDR counterparts for different beamforming designs.
\end{abstract}
\begin{IEEEkeywords}
Quadratically constrained quadratic program, semidefinite relaxation, transmit beamforming, reconfigurable intelligent surfaces.
\end{IEEEkeywords}

\IEEEpeerreviewmaketitle
\section{Introduction}
\lettrine[findent=1pt]{{\textbf{Q}}}{uadratically} constrained quadratic programs (QCQPs)  are typical optimization problems having objective functions and constraints as quadratic functions of optimization variable vectors \cite{Zhi,Huang2010new}. Indeed, QCQPs capture numerous classic research problems in signal processing and communications such as multiple-input, multiple-output detection, multi-user detection, magnetic resonance imaging, transmit beamforming, etc. \cite{Zhi}. Unfortunately, most of the QCQPs are known as non-convex problems which are generally NP-hard and cannot be solved in polynomial time \cite{Zhi,Huang2010new}. To tackle these problems, a non-convex QCQP is first transformed into an equivalent semidefinite programming by introducing a new positive semidefinite optimization variable matrix as the product of the original optimization variable vector and its hermitian. However, this equivalent transformation also introduces a non-convex rank-one constraint on the newly introduced optimization variable matrix, i.e., the matrix has only one linearly independent column/row. By dropping the rank-one constraint, the transformed problem becomes a convex one and can be effectively solved by numerical optimization packages, e.g., CVX \cite{Boyd}. This technique is known as semidefinite relaxation (SDR) and the dropped rank-one constraint problem is referred to as the SDR counterpart. If solving the SDR counterpart results in a rank-one optimal matrix, then the original QCQP and its SDR counterpart are equivalent, i.e., the optimal solution to the SDR counterpart is also the optimal solution to the original QCQP. Therefore, the optimal vector for the original QCQP problem can be attained as the product of the eigenvector of the rank-one optimal matrix of the SDR counterpart and the square root of its eigenvalue. Otherwise,  some rank-one approximations or the Gaussian randomize procedure \cite{Zhi} can be adopted to extract an approximated solution to QCQP from the optimal solution of the SDR counterpart. Indeed, extracting the approximated solution requires further computational resources and the obtained solution is generally suboptimal.

The first work adopted the SDR technique in transmit beamforming design can be dated back to the late 90's in \cite{Mats}, where the authors minimized the total transmit power subject to guaranteeing the minimum required signal-to-interference-plus-noise ratios (SINRs) for all the mobile users. In particular, Bengtsson and Ottersten stated in \cite{Mats} that if the SDR counterpart of the QCQP beamforming design is feasible, then there exists at least one rank-one solution. Considering a more general class of problems, i.e., separable homogeneous QCQP, by introducing soft-shaping interference constraints and individual shaping constraints to the problem considered in \cite{Mats}, Huang and Palomar \cite{Huang2010new} derived tighten upper bounds on the rank of extreme matrices in SDPs \cite{Pataki} and confirmed their special case in \cite{Mats}. They proved that if the original separable homogeneous QCQP and its dual problem are solvable or feasible and any optimal solution of the original problem is non-trivial, i.e., containing non zero matrix component, then the original separable homogeneous QCQP has a rank-one optimal solution if there are maximum two soft-shaping interference constraints and the matrices associated with the individual shaping constraints are semidefinte. Unlike the method in \cite{Huang2010new}, Li and Ma exploited the Karush–Kuhn–Tucker (KKT) conditions and matrix rank properties to show in \cite{QiangLi} that when the data matrices satisfyies certain matrix inequality conditions, rank-one solutions for separable SDPs can be found without any dependant on the number of constraints. On the other developments, we used the Lagrange duality to prove in \cite{TuanWCL2015} and \cite{Tuanglobecom15no1} that if the corresponding SDR counterpart is feasible, then it will admit a rank-one optimal solution.

Thanks to its capable of providing accurate or even near optimal approximations  \cite{Zhi}, the SDR technique has been exploited for QCQPs representing beamforming problems for simultaneous wireless information and power transfer (SWIPT), e.g., \cite{Xu2014,KwanSWIPT2022}, physical-layer security for SWIPT, e.g., \cite{TuanCL2015,Tuanglobecom15no2}, intelligent reflecting surface (IRS), e.g., \cite{Tuan2021,KwanIRS2021,KwanIRS2022_2}, and IRS-Aided SWIPT, e.g., \cite{KwanIRS2022,Gao2022}. With respect to SWIPT scenarios, adopting an assumption that all users' channel vectors are mutually and statically independent, reference \cite{Xu2014} considered the maximization of a weighted sum-power transferred to all the energy receivers while reference \cite{KwanSWIPT2022} studied a wireless-information-transfer maximization with both linear and non-linear energy-harvesting models. It was shown in \cite{Xu2014} and  \cite{KwanSWIPT2022} when the considered channels are statistically independent with each other, with probability one, the beamforming matrices for the information receivers are rank one while those for the energy receivers have rank less or equal one. Considering physical-layer security for SWIPT scenarios, it was proved that the SDR counterpart of a QCQP proposed in  \cite{TuanCL2015}, i.e., a total-transmit-power minimization problem,  and the SDR counterpart of a QCQP introduced in \cite{Tuanglobecom15no2}, i.e., the maximization problem of the intended signal power at every information receiver and the total received power at each energy harvesting receiver, yield rank-one optimal solutions if they are feasible. Investigating an IRS scenario in \cite{KwanIRS2022_2}, the authors showed that the optimal beamforming rank-one matrix for their IRS SDR counterpart can always be obtained for any positive transmit power budget at BS. On the other hand, as for the IRS-Aided SWIPT scenarios, the authors of \cite{KwanIRS2022} indicated that the optimal solution of the information beamforming matrix of their SDR problem always satisfies the rank-one requirement for a given positive required SINR level. It is shown in \cite{Gao2022} that there always exists an optimal information beamforming matrix to the  SDR counterpart of the weighted-sum-energy-harvesting-power maximization problem satisfying rank-one condition if the SDR is feasible. Whereas, the SDR counterpart of the weighted-sum-rate maximization problem proposed in \cite{Gao2022} yields optimal information beamforming matrix with rank less or equal one if it is feasible. 

QCQP can also capture a class of robust beamforming where the input of the optimization, i.e., the channel state information (CSI) between a wireless transmitter and its mobile users, is impaired by some errors. In such cases, the imperfection of the estimated CSI is modeled as a vector with norm-bounded random error elements. To avoid handling an infinite number of constraints due to the randomness and the continuity of the error vectors, a worst-case approach adopts the S-procedure \cite{Boyd} to transform the related QCQP constraint into a robust linear matrix inequality (LMI) constraint, see e.g., \cite{Song_lou,DerrickPT,Khandaker}. As such, a robust semidefinite program (SDP) can be formed. In particular, the authors in \cite{Song_lou} analytically showed that the SDR counterpart of the transformed SDP always yields rank-one optimal solution when the channel norm-bounded value is small or the transmitter is equipped with two antennas. Also, it is shown in \cite{DerrickPT} that when the requirements in the robust SDR counterpart constraints are less stringent, the SDR counterpart has higher chance of attaining a rank-one matrix solution.  Beside, adopting similar approach as in \cite{QiangLi}, the authors of \cite{Khandaker} proved that the robust counterpart SDR of their robust secrecy beamforming problem always yields rank-one optimal solution matrices when it is feasible.

The aforementioned worst-case approach is considered as a conservative design as it requires an exceedingly large amount of system resource to prevent rarely extreme cases \cite{Feng15,Fuhui2016,zheng2016,KhandakerSep2016}. Therefore, less conservative approaches have been proposed for robust beamforming designs by tolerating the violation of the constraints with certain chances or probabilities, e.g., \cite{Kun-Yu2014,Feng15,Fuhui2016,zheng2016,KhandakerSep2016,Tuan2021}. Unfortunately, the newly introduced probabilistic constraints neither admit convexity nor have simple closed forms due to the sophisticated probability density functions. To overcome the obstacle, safe approximation techniques are applied to replace the original non-convex probabilistic constraint by a convex constraint which is known as convex approximation \cite{Kun-Yu2014}. The problem adopting the safe approximation serves as a performance lower bound as a convex subset of the original solution set is considered \cite{Ben-tal,Kun-Yu2014}. To this end, three safe approximation methods have been developed in \cite{Kun-Yu2014}. For the first method, chance or probabilistic constraints of the original optimization problem are approximated by LMIs based on the assumption of a norm-bounded  of the error vectors and exploiting the S-procedure. As for the other two methods, large deviation inequalities for complex Gaussian quadratic forms are adopted to safely approximate a chance constraint by a set of LMI and second-order-cone (SOC) constraints leading to the formulation of rank-constrained SDP. The attempts towards the related rank-one issue of SDR containing SOC constraints were firstly accounted  for \cite{zheng2016} and latter for \cite{KhandakerSep2016}. In these works, some inequalities have been exploited to transform a SOC constraints into an LMI constraint. The resulting transformed SDRs are safe approximations, i.e., every feasible solution to the approximated problem is also feasible for the corresponding original problem \cite{Ben-tal,Kun-Yu2014}, and they are shown to yield rank-one solutions. However, as a result of the employed transformations, the sizes of the feasible regions of the transformed SDRs are reduced, in comparison with the original problem, which may lead to infeasibility, i.e., a feasible solution to the original problem may be  infeasible for the approximated problem. In \cite{TuanTGCN2017}, we proposed a novel approach to transform an SOC constraint into an LMI constraint without loss of optimality as we do not reduce the feasibility region of the transformed SDRs. We then proved that the transformed SDRs are tight, i.e., yielding rank-one optimal solutions. The finding, however, only captures the  probabilistic optimization problem considered in \cite{TuanTGCN2017}. 

The effectiveness of solving a non-convex QCQP depends on how one can extract the optimal solution to the original QCQP from the optimal solution to the SDR counterpart. 
Having rank-one optimal solution for the SDR counterpart of a QCQP implies that the relaxation is tight, i.e., the original QCQP and its SDR are equivalent. The observation of rank-one property of the optimal solution for several SDRs in literature indicates the existence of a general optimization framework. If a QCQP  can be transformed into a SDR problem subsumed under a general framework  admitting rank-one optimal solution, then the relaxation is tight and its optimal solution can be efficiently obtained via solving the corresponding SDR problem. Unfortunately, such framework has not been formulated and investigated yet. Furthermore, as the framework captures various types of constraints, proving the rank-one optimal solution to the general framework poses a challenging task, i.e., any aforementioned techniques cannot straightforwardly be utilized. Therefore, the novelty of this work is to formulate and prove the rank-one optimal solution of the general framework.  Particularly, the contributions of this paper are summarized as follows. 
\begin{itemize}
    \item  This paper proposes a general optimization problem framework including a mix of LMI and SOC constraints which can serve as the SDR counterpart of several beamforming QCQPs. The SDR counterparts considered in previous works in \cite{Mats,Huang2010new,QiangLi,TuanWCL2015,Tuanglobecom15no1,TuanCL2015,Tuanglobecom15no2,Song_lou,DerrickPT,Khandaker,zheng2016,KhandakerSep2016,Tuan2021,TuanTGCN2017} are special cases of our proposed optimization problem. 
    \item This paper proves that the proposed optimization problem framework always yields rank-one optimal solution if it is feasible and its optimal solution is non-trivial. The technique developed in this paper is the generalization of our previous works in \cite{TuanWCL2015,Tuanglobecom15no1}, and \cite{TuanTGCN2017}.
    \item This paper studies the transmit beamforming with four illustrative examples to highlight possible applications of the proposed optimization framework in developing SDR counterparts for QCQP transmit beamforming designs with both perfect and imperfect CSI. The applications of the framework are beyond these four examples which are only given to showcase how different types of constraints can be handled by the framework. 
\end{itemize}

\emph{\textbf{Notation}:}
Lower and upper case letter $y$ and $Y$: a scalar; bold lower case letter $\mathbf{y}$: a column vector; bold upper case letter $\mathbf{Y}$: a matrix; $\left\|\cdot\right\|$: the Euclidean norm; $\left\|\cdot\right\|_F$: the Frobenius norm; $(\cdot)^T$: the transpose operator; $(\cdot)^H$: the complex conjugate transpose operator; $\textrm{Tr}\left(\cdot\right)$: the trace operator; $\textrm{Pr}\left(\cdot\right)$: the probability of an event; $\mathbf{Y}\succeq \mathbf{0}$: $\mathbf{Y}$ is positive semidefinite; $\mathbf{y}\succcurlyeq \mathbf{0}$: all elements of vector $\mathbf{y}$ are non-negative; $\mathbf{I}_x$: an $x \times x$ identity matrix; $\mathbf{0}_{A \times 1}$: an $A \times 1$ vector of all zero elements; $\mathbf{0}_{A \times B}$: an $A \times B$ matrix of all zero elements; $\textrm{Re}\{\cdot\}$: the real part of a complex number; $\textrm{Eig}_{\textrm{max}}\left(\mathbf{Y}\right)$: the maximum eigenvalue of $\mathbf{Y}$; $s^+(\mathbf{Y}): \ \textrm{max} \{\textrm{Eig}_{\textrm{max}}(\mathbf{Y}),0\}$; $\textrm{vec}\left(\mathbf{Y}\right)$: stacking all the entries of $\mathbf{Y}$ into a column vector; $\mathbb{R}$: the set of all real scalars ;  $\mathbb{C}^{M\times 1}$: the set of all $M\times 1$ vectors with complex elements; $\mathbb{H}^{M\times M}$: the set of all $M\times M$ Hermitian matrices; $y\sim\mathcal{CN}(0,\sigma^2)$: $y$ is a zero-mean circularly symmetric complex Gaussian random variable with variance $\sigma^2$; $\mathbf{y}\sim\mathcal{CN}(\mathbf{0},\mathbf{Y})$: $\mathbf{y}$ is a zero-mean circularly symmetric complex Gaussian random vector with covariance matrix $\mathbf{Y}$;  $\mathbf{Y}^{1/2}$: the square root of matrix $\mathbf{Y}$; $\textrm{diag}\left( \mathbf{y}\right)$: a diagonal matrix whose diagonal elements are the entries of vector $\mathbf{y}$; and finally $\textrm{diag}\left( \mathbf{Y}\right)$: a vector whose entries are the diagonal elements of matrix $\mathbf{Y}$.
\section{Proposed Optimization Framework}
\subsection{Rank-one Optimization Framework}
Consider the following optimization problem framework with a mix of LMI and SOC constraints:
\begin{equation}
\begin{aligned}\label{rootSDPnew}
& \displaystyle \min_{\{\mathbf{W}_{i}\}\in \mathbb{H}^{M\times M},\ \alpha_i\geq0,\  \varrho_i,\ f_i\geq0} & &
\sum_{i=1}^U \textrm{Tr}\left(\mathbf{A}_{i}\mathbf{W}_{i}\right) \\
& \text{s.\ t.}: & & C1,  \forall i \in \{1,\cdots,L_1\},\\
&&& 
 C2,  \forall i \in \{1,\cdots,U\},\\
&&& C3, \  \forall i  \in \{1,\cdots,L_3\},\\
&&&  C4,  \ \forall i  \in \{1,\cdots,L_4\},\\
&&& C5,\   \forall i  \in \{1,\cdots,L_5\},\\
&&& C6, \ \forall i \in \{1,\cdots,U\},
\end{aligned}
\end{equation}
where $\{\mathbf{W}_{i}\}=\{\mathbf{W}_{1},\mathbf{W}_{2},\cdots,\mathbf{W}_{U}\}$, 
\begin{eqnarray}
C1&:& \ a_i \textrm{Tr}\left(\mathbf{X}_{i,i}\mathbf{W}_i\right) +\sum_{j=1}^U b_j\textrm{Tr}\left(\mathbf{X}_{i,j}\mathbf{W}b_j\right)+c_i\geq 0,\\
C2&:& m_i\textrm{Tr}\left(\mathbf{M}_{i}\mathbf{W}_i\right) +p_i\geq 0,\\
C3&:& \ \begin{bmatrix} \mathbf{Y}_i\mathbf{B}_i\mathbf{Y}_i+\alpha_i\mathbf{I}_M&\mathbf{Y}_i\mathbf{B}_i\mathbf{y}_i\\\ \mathbf{y}_i^H\mathbf{B}_i\mathbf{Y}_i&\mathbf{y}_i^H\mathbf{B}_i\mathbf{y}_i+
d_i
\end{bmatrix}\succeq \mathbf{0}, \\
C4&:&\ \left \Vert\begin{bmatrix}
e_i\mathbf{Z}_i\mathbf{C}_i
\mathbf{z}_i \\
\textrm{vec}\left( \mathbf{Z}_i\mathbf{C}_i\mathbf{Z}_i\right)
\end{bmatrix}
\right \Vert\leq \varrho_i,\\
C5&:& \ f_i\mathbf{I}_M+v_i\mathbf{D}_i\sum_{k=1}^N\boldsymbol\Lambda_k\mathbf{E}_i\boldsymbol\Psi_k\widetilde{\mathbf{D}}_i\succeq \mathbf{0},\\
C6&:& \ \mathbf{W}_{i} \succeq \mathbf{0},
\end{eqnarray}
$L_1,L_3,L_4,L_5$, and $U$ are the number of constraints with $\{L_1,L_3,L_4,L_5\}\geq U$; $\{\mathbf{W}_{i}\}$, $\{\alpha_i\}$, $\{\varrho_i\}$, and $\{f_i\}$ are, respectively, the sets of $U$, $L_3$, $L_4$, and $L_5$ numbers of optimization variables; $M$ is the order/size/dimension of the variable square Hermitian matrix $\mathbf{W}_{i}$. The optimization data, i.e., given design parameters, are as follows: \{$\mathbf{A}_{i},\mathbf{X}_{i,i}, \mathbf{X}_{i,j},\boldsymbol\Lambda_k,\boldsymbol\Psi_k\} \in \mathbb{C}^{M\times M}$, \{$\mathbf{Y}_{i},\mathbf{Z}_{i},\mathbf{D}_{i},\widetilde{\mathbf{D}}_i\}\in \mathbb{H}^{M\times M} $, $\{\mathbf{y}_i,\mathbf{z}_i\}\in \mathbb{C}^{M\times 1}$, and $\{a_i$, $b_i$, $c_i$, $d_i$, $e_i$, $g_i$, $v_i$, $m_i$, $p_i$, $h_j$, $\tilde{g}_i$, $\tilde{h}_j$,  $\bar{g}_i$, $\bar{h}_j\}\in \mathbb{R}$. In \eqref{rootSDPnew}, $\mathbf{B}_i$, $\mathbf{C}_i$ and $\mathbf{E}_i$ are affine functions of $\{\mathbf{W}_i\}$, i.e., 
\begin{eqnarray}
    \mathbf{B}_i&=&g_i\mathbf{W}_i+\sum_{j=1}^U h_j\mathbf{W}_j, \\
    \mathbf{C}_i&=&\tilde{g}_i\mathbf{W}_i+\sum_{j=1}^U \tilde{h}_j\mathbf{W}_j,\\  \mathbf{E}_i&=&\bar{g}_i\mathbf{W}_i+\sum_{j=1}^U \bar{h}_j\mathbf{W}_j. 
\end{eqnarray}
Finally, $a_i=0$, $g_i=0$, $\tilde{g}_i=0$, and $\bar{g}_i=0$,  $\forall i> U$.

Exploiting the Schur complement with some mathematical manipulations, one can rewrite C4 as:
\begin{equation}
    C4(a): \begin{bmatrix}\varrho_i \mathbf{I}_{M^2+M} & \begin{bmatrix}
e_i\mathbf{Z}_i\mathbf{C}_i
\mathbf{z}_i \\
\textrm{vec}\left( \mathbf{Z}_i\mathbf{C}_i\mathbf{Z}_i\right)
\end{bmatrix}\\
\begin{bmatrix}
e_i\mathbf{Z}_i\mathbf{C}_i
\mathbf{z}_i \\
\textrm{vec}\left( \mathbf{Z}_i\mathbf{C}_i\mathbf{Z}_i\right)
\end{bmatrix}^H&\varrho_i\end{bmatrix}
\succeq \mathbf{0}.
\end{equation}
Hence, the proposed optimization problem framework \eqref{rootSDPnew} can be equivalently written as:

\begin{equation}
\begin{aligned}\label{rootSDP}
& \displaystyle \min_{\{\mathbf{W}_{i}\}\in \mathbb{H}^{M\times M},\ \alpha_i\geq0,\  \varrho_i,\ f_i\geq0} & &
\sum_{i=1}^U \textrm{Tr}\left(\mathbf{A}_{i}\mathbf{W}_{i}\right) \\
& \text{s.\ t.}\ & & C1,  \forall i \in \{1,\cdots,L_1\},\\
&&& C2,  \forall i \in \{1,\cdots,U\},\\
&&& C3, \  \forall i  \in \{1,\cdots,L_3\},\\
&&& C4(a),  \ \forall i  \in \{1,\cdots,L_4\},\\
&&& C5,\   \forall i  \in \{1,\cdots,L_5\},\\
&&& C6, \ \forall i \in \{1,\cdots,U\}.
\end{aligned}
\end{equation}

We then introduce the following theorem.
\begin{theorem}\label{theo1}
If problem \eqref{rootSDP} is feasible and each of its optimal solution $\mathbf{W}_{i}^{\star}\neq \mathbf{0}$, $\forall i$, then $\mathbf{W}_{i}^{\star}$, $\forall i$,  are rank-one matrices.\footnote{Since \eqref{rootSDP} is convex, $\mathbf{W}_{i}^{\star}$ is a unique solution in the considered feasible region.}
\end{theorem}
\begin{proof}
Please refer to Appendix \ref{apen1}.
\end{proof}

With a suitable selection of $\mathbf{A}_i$, e.g., $\mathbf{A}_i=-\mathbf{I}_M$, the minimization can be turned into a maximization. For instance, the proposed framework can capture the weighted sum-power transfer maximization in \cite{Xu2014,Gao2022}, the wireless-information transfer efficiency maximization in \cite{KwanSWIPT2022}, the intended signal power maximization in \cite{Tuanglobecom15no2}, etc. 

For a sum rate maximization problem, which can be considered as a fractional QCQP, one can follow similar steps as in \cite[Section IV]{Gao2022} to cast it in the form of the proposed framework. On the other development, a novel approach was introduced in \cite{Nao19} to approximate the sum rate maximization problem by a convex SOC form which does not require SDR technique.\footnote{A similar approximation technique was adopted in \cite{Nao21} to find solution for a energy efficiency maximization problem.} Recently, elegant approaches have been introduced in \cite{Luo21} and \cite{Luo22} to tackle factional QCQPs without the need of SDR technique. As a result, the proposed approaches are capable of  solving sum rate optimization problems with both perfect and imperfect CSI. It has been shown in \cite{Luo21} and \cite{Luo22} that their proposed approaches offer better performances than their SDR counterparts do. 

Note that none of the SDR counterparts of the transmit beamforming approaches in the literature considers all the constraints presented in the optimization framework. Constraints $C1$, $C2$, $C4/C4(a)$, and $C6$ usually appear on beamforming problems with perfect CSI, see e.g., \cite{Mats,Huang2010new,Pataki,QiangLi,TuanWCL2015,Tuanglobecom15no1,TuanCL2015,Tuanglobecom15no2}. As for imperfect CSI scenarios, constraints $C3$ and $C6$ are adopted in robust conservative beamforming approaches, such as \cite{Song_lou,DerrickPT,Khandaker}, whereas constraints  $C4/C4(a)$, $C5$, and $C6$ are exploited in probabilistic beamforming approaches, e.g., \cite{Kun-Yu2014,Feng15,Fuhui2016,zheng2016,KhandakerSep2016,Tuan2021}.

When $\{L_1,L_3,L_4,L_5\} \leq U$, $a_i \neq 0$, $g_i\neq 0$, $\tilde{g}_i\neq 0$, and $\bar{g}_i\neq 0$. Hence, constraints $C1$, $C3$, $C4/C4(a)$, and $C5$ are exploited to represent the SINR constraints. When $\{L_1,L_3,L_4,L_5\} > U$, $a_i=0$, $g_i=0$, $\tilde{g}_i=0$, and $\bar{g}_i=0$. Therefore, constraints $C1$, $C3$, $C4/C4(a)$, and $C5$ can represent the soft-shaping interference constraint, e.g., as in \cite{Huang2010new,TuanWCL2015}, or the energy
transfer constraint, e.g., as in \cite{Khandaker,TuanTGCN2017}. Constraint $C2$ can capture the individual shaping constraint of the SDR in \cite{Huang2010new}.

Let us define a sub-class optimization problem of the general framework as an optimization problem including an objective function as in  \eqref{rootSDP}, constraint $C6$ and at least one constraint selected from $\{C1,C2,C3,C4(a),C5\}$. As the result of Theorem ~\ref{theo1}, we have the following corollary.
\begin{cor}\label{cor01}
Sub-class optimization problems  of the general framework in \eqref{rootSDP} yield rank-one-matrix optimal solutions if they are feasible and there is no trivial solution amongst their optimal solutions.
\end{cor}
\begin{proof}
The proof is the simplified version of that presented in Appendix \ref{apen1}.
\end{proof}
\begin{remark}
The rank-one results in Theorem~\ref{theo1} and Corollary~\ref{cor01} are based on an assumption that the problem \eqref{rootSDP} or its sub-class problem is feasible and there is no trivial solution. However, the conditions for such assumption being held, e.g., the relationship between the number of constraints and the number of variables, or the input data of the problem, or the channel estimation errors, are out of the scope of this paper. Theorem~\ref{theo1} and Corollary~\ref{cor01} indicate that the optimization framework in \eqref{rootSDP} can serve as tight SDR counterparts for its corresponding QCQPs. In other words, only those QCQPs being able to convert into the general framework \eqref{rootSDP} or it sub-class optimization problems have tight SDRs.
\end{remark}
\subsection{Complexity Analysis}
Since the proposed optimization framework \eqref{rootSDP} is convex, a standard interior point method (IPM) can be exploited to find the optimal solution. To that end, we analyse the complexity of solving \eqref{rootSDP} in a worst-case run time of the IPM. We start by introducing the following definition.
\begin{definition}
At a given $\varepsilon>0$, the set of $\{\mathbf{W}_i^{\varepsilon} \}$ is an $\varepsilon$-solution to problem \eqref{rootSDP}, i.e., an acceptable solution with the accuracy of $\varepsilon$, if
\begin{equation}
    \sum_{i=1}^U\textrm{Tr}\left(\mathbf{A}_i\mathbf{A}_i^{\varepsilon}\right)\leq \displaystyle \min_{\mathbf{W}_i\in \mathbb{H}^{M\times M}}  \sum_{i=1}^U\textrm{Tr}\left(\mathbf{A}_i\mathbf{W}_i\right) +\varepsilon.
\end{equation}
\end{definition}
It can be seen that the number of decision variables of \eqref{rootSDP} is $M^2$. The complexity of \eqref{rootSDP} is described in the following lemma.
\begin{lemma}\label{lemSDPcog}
The computational complexity to attain $\varepsilon$-solution to \eqref{rootSDP} is on the order of:
\begin{align}\label{Comp_SDPcog}
    \ln{\left(\varepsilon^{-1}\right)}\sqrt{\beta\Big(\mathcal{M}\Big)}&\Big[ C_{\text{form}}+C_{\text{fact}}\Big],
\end{align}
where $\beta(\mathcal{M})=L_1+U+L_3+L_4+\Big(L_3+L_4+L_5+U\Big)M+L_4M^2$, $C_{\text{form}}=M^2\Big[ L_1+U+\Big(L_5+U\Big)M^3+L_3\Big(M+1\Big)^3+L_4\Big(M^2+M+1\Big)^3\Big]+M^4\Big[ L_1+U+L_3\Big( M+1\Big)^2+L_4\Big( M^2+M+1\Big)^2+\Big( L_5+U\Big)M^2\Big]$, and $C_{\text{fact}}=6M^6$.
\end{lemma}
\begin{proof}
We sketch some main steps to arrive at the lemma due to space limitation. It can be observed that \eqref{rootSDP} has $(L_1+U)$ linear-matrix-inequality (LMI) constraints of size 1, $(L_5+U)$ LMI constraints of size $M$, $L_3$ LMI constraints of size $(M+1)$, and $L_4$ LMI constraints of size $(M^2+M+1)$. One can follow the same steps as in \cite[Section V-A]{Kun-Yu2014} to derive the following facts: (i) the iteration complexity is on the order of $\ln{\left(\varepsilon^{-1}\right)}\sqrt{\beta\Big(\mathcal{M}\Big)}$, and (ii)  the per-iteration complexity is on the order of $C_{\text{form}}+C_{\text{fact}}$.
\end{proof}

In the following section, we present how the proposed framework \eqref{rootSDP} can be adopted to solve transmit beamforming QCQPs.
\section{Applications to Transmit Beamforming}
As discussed in the previous section, the constraints in the optimization framework \eqref{rootSDP} can well represent several constraint types of the SDR counterparts of most of the QCQP transmit beamforming designs in the literature. In this section, we consider a downlink beamforming scenario  and present four illustrative examples with either perfect or imperfect CSI. Those examples show how these types of the constraints in the proposed optimization framework can be adopted to obtain the globally optimal solution of QCQP transmit beamforming designs.
\subsection{Transmit Beamforming with Perfect CSI}\label{TxPerfectCSI}
Consider a cellular system consisting of a base station (BS) serving
$U$ mobile users. We assume that
the BS is equipped with $M$ antenna elements and each mobile user
has a single antenna. The received signal at user $i$, $ i
\in \{1, \cdots, U\}$, is:
\begin{eqnarray}\label{signal1}
y_{i}=\mathbf{h}^H_{i}\mathbf{w}_{i}s_{i}+\sum_{j=1,j
\neq i}^{U}\mathbf{h}^H_{i}\mathbf{w}_{j}s_{j}+n_{i},
\end{eqnarray}
where $\mathbf{h}^H_{i}\in\mathbb{C}^{1 \times M}$ is the channel coefficient
between user $i$ and the BS; $\mathbf{w}_{i}\in\mathbb{C}^{M \times 1}$  and
$s_{i}\sim\mathcal{CN}(0,1)$ are the beamforming vector and the data symbol associated to
 user $i$, respectively; and $n_{i}\sim\mathcal{CN}(0,\sigma^2_i)$ is a zero mean circularly
symmetric complex Gaussian noise with variance $\sigma^2_i$, at user $i$. We express the SINR at any user $i$ as:
\begin{equation}
\text{SINR}_{i}=\frac{\mathbf{w}_{i}^H \mathbf{h}_{i}\mathbf{h}^H_{i}\mathbf{w}_{i}}{\sum_{j=1,j \neq i}^{U}\mathbf{w}_{j}^H\mathbf{h}_{i}\mathbf{h}^H_{i}\mathbf{w}_{j}+\sigma^2_i}, \ \forall i.\label{second}
\end{equation}

Our objective is to design downlink beamforming
vectors for the mobile users that minimize the BS's transmit power while maintaining the required SINR level for each user. The optimization problem to design the beamforming vectors can be stated as:
\begin{equation}
\begin{aligned}\label{cog_perfectCSI}
& \displaystyle \min_{\mathbf{w}_{i}} & &
\sum_{i=1}^U \|\mathbf{w}_{i}\|^2\\
& \text{s.\ t.}\ & &\frac{\mathbf{w}^H_{i}\mathbf{h}_{i}\mathbf{h}^H_{i}\mathbf{w}_{i}}
{\sum_{j =1,j \neq i}^U\mathbf{w}^H_{j}\mathbf{h}_{i}\mathbf{h}^H_{i}
\mathbf{w}_{j}+\sigma^2_i}\geq \gamma_{i}, \ \forall i \in \{1,\cdots,U\},
\end{aligned}
\end{equation}
where $\gamma_i$ is the user $i$ target SINR level. 

Due to the SINR constraint, problem \eqref{cog_perfectCSI} is non-convex with respect to $\mathbf{w}_i$. In this paper, we are interested in casting problem \eqref{cog_perfectCSI} in a QCQP form.\footnote{Problem \eqref{cog_perfectCSI} can also be transformed into a second-order-cone programming which is not a focus of this paper. Interested readers are referred to \cite[Section IV.B]{Ami} for the detailed transformation.} Let us introduce a new optimization variable $\mathbf{W}_i=\mathbf{w}_i\mathbf{w}_i^H$ where $\mathbf{W}_{i}\succeq \mathbf{0}$, $\mathbf{W}_{i}\in \mathbb{H}^{M\times M}$, and $\mathbf{W}_{i}$ is a rank-one matrix. Utilizing the identities $\mathbf{x}^H\mathbf{Y}\mathbf{x}=\textrm{Tr}(\mathbf{Y}\mathbf{x}\mathbf{x}^H)$ and $\|\mathbf{x}\|^2=\textrm{Tr}(\mathbf{x}\mathbf{x}^H)$ with some mathematical manipulations, one can equivalently rewrite \eqref{cog_perfectCSI} as:
\begin{equation}
\begin{aligned}\label{cog_perfect_1}
& \displaystyle \min_{\mathbf{W}_i\in \mathbb{H}^{M\times M}} & & \sum_{i=1}^U\textrm{Tr}\left(\mathbf{W}_i\right)\\
& \text{s.\ t.}\ & & \left(1+\frac{1}{\gamma_i} \right) \textrm{Tr}\left(\mathbf{h}_{i}\mathbf{h}^H_{i}\mathbf{W}_i\right)-\sum_{j=1}^U\textrm{Tr}
\left(\mathbf{h}_{i}\mathbf{h}^H_{i}\mathbf{W}_{j}\right)-\sigma_i^2\geq0,\\
&&& \ \ \ \ \ \ \ \ \ \ \ \ \ \ \ \ \ \ \ \ \ \ \ \ \ \ \  \ \ \ \ \ \ \ \ \ \ \ \ \  \forall i \in \{1,\cdots,U\}, \\ 
&&& \mathbf{W}_i\succeq \mathbf{0},\ \forall i \in \{1,\cdots,U\},\\
&&& \textrm{rank}\left( \mathbf{W}_i\right)=1, \ \forall i \in \{1,\cdots,U\}.
\end{aligned}
\end{equation}

In the following we show that the SINR constraint in \eqref{cog_perfect_1} can be expressed the form of $C1$ in  \eqref{rootSDP}. To that end, we map the notations used in \eqref{cog_perfect_1} to those used in \eqref{rootSDP} as follows. Let us denote $a_i=\left(1+\frac{1}{\gamma_i} \right)$, $b_j=-1$, $\mathbf{X}_{i,i}=\mathbf{X}_{i,j}=\mathbf{h}_{i}\mathbf{h}^H_{i}$, and $c_i=-\sigma_i^2$.  Dropping the rank-one constraint on $\mathbf{W}_i$, utilizing those mappings, and letting $\mathbf{A}_i=\mathbf{I}_M$, \eqref{cog_perfect_1} can be rewritten as:
\begin{equation}
\begin{aligned}\label{cog_perfect_2}
& \displaystyle \min_{\mathbf{W}_i\in \mathbb{H}^{M\times M}} & & \sum_{i=1}^U\textrm{Tr}\left(\mathbf{A}_i\mathbf{W}_i\right)\\
& \text{s.\ t.}\ & & a_i\textrm{Tr}\left(\mathbf{X}_{i,i}\mathbf{W}_i\right)+\sum_{j=1}^U b_j\textrm{Tr}
\left(\mathbf{X}_{i,j}\mathbf{W}_{j}\right)+c_i\geq0, \\
&&& \ \ \ \ \ \ \ \ \ \ \ \ \ \ \ \ \ \ \ \ \ \ \ \ \ \ \ \ \ \   \forall i \in \{1,\cdots,U\}, \\ 
 &&& \mathbf{W}_i\succeq  \mathbf{0},\ \forall i \in \{1,\cdots,U\}.
\end{aligned}
\end{equation}

It is clear that \eqref{cog_perfect_2} is a sub-class optimization problem of the general framework \eqref{rootSDP}, i.e., containing $C_1$ and $C_6$ with $L_1=U$. Hence, according to Corollary~\ref{cor01}, \eqref{cog_perfect_2}  yields  rank-one optimal solution if it is feasible and its optimal solution does not contain any trivial solution. Therefore, the optimal solution of \eqref{cog_perfect_2} is also the optimal solution of \eqref{cog_perfect_1}. In other words, the SDR counterpart \eqref{cog_perfect_2} is equivalent to the original QCQP \eqref{cog_perfect_1}.

Keeping $C_1$ and $C_6$ in the general framework \eqref{rootSDP} with $L_1=U$, one can derive the complexity of \eqref{cog_perfect_2} from Lemma~\ref{lemSDPcog} as follows.

\begin{cor}
    The computational complexity to attain $\varepsilon$-solution to \eqref{cog_perfect_2}  is on the order of:
\begin{align}
\ln{\left(\varepsilon^{-1}\right)}\sqrt{\beta_1\Big(\mathcal{M}\Big)}\Big[ C_{\text{form},1}+C_{\text{fact},1}\Big],\label{complex01}
\end{align}
where $\beta_1(\mathcal{M})=U(M+1)$, $C_{\text{form},1}=M^2\Big[ U(M^3+1)\Big]+M^4\Big[U(M^2+1)\Big]$, and $C_{\text{fact},1}=2M^6$.
\end{cor}
\subsection{Transmit Beamforming with Imperfect CSI}
Consider the same system described in the previous subsection. However, we assume that there exist errors in the estimation of the channel state information.  Hence, the true channel coefficient between the BS and user $i$ is modeled as $\widehat{\mathbf{h}}_{i}^H+\mathbf{H}_{i}^{1/2}\mathbf{e}_{i}^H$ where $\widehat{\mathbf{h}}_{i}$ is the estimate channel coefficient, $\mathbf{H}_{i} \succeq \mathbf{0}$ and $\mathbf{e}_{i}\sim\mathcal{CN}(\mathbf{0},\mathbf{I}_M)$ represent its estimation error. We further assume that the error vector $\mathbf{e}_{i}$ is confined to the complex spherical sets $ \{\mathbf{e}_{i} \in \mathbb{C}^{M\times 1}\ | \ \|\mathbf{e}_{i}\|^2\leq r^2\}$ having $M$ dimensions and radius $r$. Our optimization for imperfect CSI is posed as:
\begin{equation}
\begin{aligned}\label{cog_imperfectCSI}
& \displaystyle \min_{\mathbf{w}_{i}} & &
\sum_{i=1}^U \|\mathbf{w}_{i}\|^2\\
& \text{s.\ t.}\ & &\frac{\mathbf{w}^H_{i}\left(\widehat{\mathbf{h}}_{i}+\mathbf{H}_{i}^{1/2}\mathbf{e}_{i}\right)
\left(\widehat{\mathbf{h}}_{i}+\mathbf{H}_{i}^{1/2}\mathbf{e}_{i}\right)^H\mathbf{w}_{i}}
{\sum_{j =1,j \neq i}^U\mathbf{w}^H_{j}\left(\widehat{\mathbf{h}}_{i}+\mathbf{H}_{i}^{1/2}\mathbf{e}_{i}\right)
\left(\widehat{\mathbf{h}}_{i}+\mathbf{H}_{i}^{1/2}\mathbf{e}_{i}\right)^H
\mathbf{w}_{j}+\sigma^2_i}\geq \gamma_{i},\\
&&& \ \ \ \ \ \ \ \ \ \ \ \ \ \ \ \ \ \ \ \ \ \ \ \ \ \ \ \ \ \ \ \  \ \ \ \ \ \ \ \ \ \ \forall i \in \{1,\cdots,U\}, \\
&&& \|\mathbf{e}_{i}\|^2\leq r^2,\ \forall i \in \{1,\cdots,U\}.
\end{aligned}
\end{equation}

Introducing a new optimization variable $\mathbf{W}_i=\mathbf{w}_i\mathbf{w}_i^H$ where $\mathbf{W}_{i}\succeq \mathbf{0}$, $\mathbf{W}_{i}\in \mathbb{H}^{M\times M}$, and $\mathbf{W}_{i}$ is a rank-one matrix, we rewrite the SINR constraint in \eqref{cog_imperfectCSI} as an affine function of $\mathbf{e}_{i}$:
\begin{eqnarray}\label{sinr_event}
f_i(\mathbf{e}_{i})&\triangleq&\mathbf{e}_{i}^H\mathbf{H}_{i}^{1/2}\mathbf{\widetilde{B}}_i
\mathbf{H}_{i}^{1/2}\mathbf{e}_{i}+2 \textrm{Re} \{\mathbf{e}_{i}^H\mathbf{H}_{i}^{1/2}\mathbf{\widetilde{B}}_{i}\widehat{\mathbf{h}}_{i}\} +\widehat{\mathbf{h}}_{i}^H\mathbf{\widetilde{B}}_i\widehat{\mathbf{h}}_{i}-\sigma_i^2 \nonumber \\&\geq& 0,
\label{co_sinr2}
\end{eqnarray}
where 
\begin{equation}
    \mathbf{\widetilde{B}}_i=\left( 1+\frac{1}{\gamma_i}\right)\mathbf{W}_i-\sum_{j=1}^U\mathbf{W}_j. \label{B_i}
\end{equation}
We now equivalently rewrite \eqref{cog_imperfectCSI} as:
\begin{equation}
\begin{aligned}\label{cog_imperfectCSI2}
& \displaystyle \min_{\mathbf{W}_{i}\in \mathbb{H}^{M \times M}} & &
\sum_{i=1}^U\textrm{Tr}\left(\mathbf{W}_i\right)\\
& \text{s.\ t.}\ & &f_i(\mathbf{e}_{i})\geq 0, \ \forall i \in \{1,\cdots,U\},
\\
&&& \mathbf{e}_{i}^H\mathbf{I}\mathbf{e}_{i} \leq r^2,\ \forall i \in \{1,\cdots,U\},\\
&&&\mathbf{W}_i\succeq 0,\ \textrm{rank}\left( \mathbf{W}_i\right)=1, \ \forall i \in \{1,\cdots,U\}.
\end{aligned}
\end{equation}
The number of constraints in \eqref{cog_imperfectCSI2} is infinite\footnote{ Problem \eqref{cog_imperfectCSI2} is a semi-infinite optimization problem, i.e., an optimization problem with a finite number of variables and an infinite number of constraints.} due to the randomness and continuous of the error vector $\mathbf{e}_{i}$. To proceed, we introduce the following lemma.
\begin{lemma}
[S-Procedure\cite{Boyd_convex}] Let
$m_n(\mathbf{x})=\mathbf{x}^H\mathbf{Y}_n\mathbf{x}+2\textrm{Re}
\{\mathbf{x}^H\mathbf{y}_n\}+c_n,\ n\in\{1,2\}$,
where $\mathbf{Y}_n\in \mathbb{H}^{M\times M}$, $\mathbf{y}_n\in \mathbb{C}^{M\times 1}$, and $c_n\in \mathbb{R}$. If there exists an $\check{\mathbf{x}}$ such that $m_n(\check{\mathbf{x}})<0$, then $\forall \mathbf{x} \in \mathbb{C}^{M\times 1}$, the following statements are equivalent:
\begin{enumerate}
\item $m_1(\mathbf{x})\geq 0$ and $m_2(\mathbf{x})\leq 0$ are satisfied $\forall \mathbf{x} \in \mathbb{C}^{M\times 1}$.
\item There exists a $\beta \geq 0$ such that $
\begin{bmatrix} \mathbf{Y}_1&\mathbf{y}_1\\\ \mathbf{y}_1^H&c_1\end{bmatrix}+\beta
\begin{bmatrix} \mathbf{Y}_2&\mathbf{y}_2\\\ \mathbf{y}_2^H& c_2\end{bmatrix}\succeq \mathbf{0}$.
\end{enumerate}
\end{lemma}
Exploiting Lemma 1, one can transform the optimization problem \eqref{cog_imperfectCSI2} into an equivalent standard convex SDP form  as:
\begin{equation}
\begin{aligned}\label{prob_sdp3}
& \displaystyle \min_{\{\mathbf{W}_{i}\}\in \mathbb{H}^{M \times M},\ \widetilde{\beta}_i\geq 0} & & \sum_{i=1}^U\textrm{Tr}\left(\mathbf{W}_{i}\right)\\
& \text{s.\ t.}\ & & \begin{bmatrix} \mathbf{H}_{i}^{1/2}\mathbf{\widetilde{B}}_i\mathbf{H}_{i}^{1/2}+\widetilde{\beta}_i\mathbf{I}_M&\mathbf{H}_{i}^{1/2}\mathbf{\widetilde{B}}_i\widehat{\mathbf{h}}_{i}\\\ \widehat{\mathbf{h}}_{i}^H\mathbf{\widetilde{B}}_{i}\mathbf{H}_{i}^{1/2}&\widehat{\mathbf{h}}_{i}^H\mathbf{\widetilde{B}}_i\widehat{\mathbf{h}}_{i}-\sigma_i^2-
\widetilde{\beta}_i r^2\end{bmatrix}\succeq \mathbf{0}, \\
&&& \ \ \ \ \ \ \ \ \ \ \ \ \ \ \ \ \ \ \ \ \ \ \ \ \ \ \  \ \ \forall i \in\{1,\cdots,U\},\\
 & & &
\mathbf{W}_{i}\succeq \mathbf{0},\ \textrm{rank}\left(\mathbf{W}_{i}\right)=1, \ \forall i \in \{ 1, \cdots, U\}.
\end{aligned}
\end{equation}

Note that the SINR constraint in \eqref{cog_imperfectCSI} is now cast as the first constraint in \eqref{prob_sdp3}. In the following we show that the SINR constraint in \eqref{prob_sdp3} can be expressed the form of $C3$ in  \eqref{rootSDP}. We denote $\mathbf{Y}_i=\mathbf{H}_{i}^{1/2}$, $\mathbf{B}_i=\mathbf{\widetilde{B}}_i$, $\mathbf{y}_i=\widehat{\mathbf{h}}_{i}$, $\alpha_i=\widetilde{\beta}_i$, and $d_i=-\sigma_i^2-
\widetilde{\beta}_ir^2$. Dropping the rank-one constraint on $\mathbf{W}_i$, utilizing those mappings, and letting $\mathbf{A}_i=\mathbf{I}_M$, problem \eqref{prob_sdp3} can be rewritten as:
\begin{equation}
\begin{aligned}\label{imperfectCIS_SDP}
& \displaystyle \min_{\{\mathbf{W}_{i}\}\in \mathbb{H}^{M\times M},\alpha_i\geq0} & &
\sum_{i=1}^U \textrm{Tr}\left(\mathbf{A}_{i}\mathbf{W}_{i}\right) \\
& \text{s.\ t.}\ & & \begin{bmatrix} \mathbf{Y}_i\mathbf{B}_i\mathbf{Y}_i+\alpha_i\mathbf{I}_M&\mathbf{Y}_i\mathbf{B}_i\mathbf{y}_i\\\ \mathbf{y}_i^H\mathbf{B}_i\mathbf{Y}_i&\mathbf{y}_i^H\mathbf{B}_i\mathbf{y}_i+
d_i
\end{bmatrix}\succeq \mathbf{0}, \\
&&& \ \ \ \ \ \ \ \ \ \ \ \ \ \ \ \ \ \ \ \ \ \ \  \forall i  \in \{1,\cdots,U\},\\
&&& \mathbf{W}_{i} \succeq \mathbf{0}, \ \forall i \in \{1,\cdots,U\}.
\end{aligned}
\end{equation}

It is obvious that \eqref{imperfectCIS_SDP} is a sub-class optimization problem of the general framework \eqref{rootSDP}, containing constraints $C_3$ and $C_6$ with $L_3=U$. Therefore, Corollary~\ref{cor01} indicates that \eqref{imperfectCIS_SDP} yields rank-one optimal solution when it is feasible and its optimal solution is not trivial. Hence, the optimal solution of \eqref{imperfectCIS_SDP} is the same as that of \eqref{prob_sdp3}. Therefore, the SDR counterpart \eqref{imperfectCIS_SDP} is equivalent to the original QCQP \eqref{prob_sdp3}.

Having only constraints $C_3$ and $C_6$ with $L_3=U$ in general framework \eqref{rootSDP}, one can derive the complexity of \eqref{imperfectCIS_SDP} from Lemma~\ref{lemSDPcog} as follows.

\begin{cor}
    The computational complexity to attain $\varepsilon$-solution to \eqref{imperfectCIS_SDP}  is on the order of:
\begin{align}\label{Comp_SDPcog}
    \ln{\left(\varepsilon^{-1}\right)}\sqrt{\beta_2\Big(\mathcal{M}\Big)}&\Big[ C_{\text{form},2}+C_{\text{fact},2}\Big],
\end{align}
where $\beta_2(\mathcal{M})=U(2M+1)$, $C_{\text{form},2}=M^2\Big[ U(M+1)^3+UM^3\Big]+M^4\Big[U(M+1)^2+UM^2\Big]$, and $C_{\text{fact},2}=2M^6$.
\end{cor}
\subsection{Chance-constraint Approach for Imperfect CSI }
In this section, we relax the constraint on the bounded-norm of the error vector to develop a probabilistic-constraint method. To that end, we define a communication outage between the BS and its users if the SINR level at the $i$-th user falls bellow a required level $\gamma_i$, which is referred to as {\it SINR outage}, see e.g., \cite{Kun-Yu2014}, \cite{TuanTGCN2017}, and the references therein. Aiming to design a power-efficient beamforming scheme, we optimize the data beamforming vector set $\{\mathbf{w}_{i}\}$ for the minimization of the total transmit power subject to probabilistic/chance constraint on SINR outages. The design is formulated as the following optimization problem:
\begin{equation}
\begin{aligned}\label{chance_ori}
& \displaystyle \min_{\{\mathbf{w}_{i}\}} & &
\sum_{i=1}^U \|\mathbf{w}_{i} \|^2\\
& \text{s.\ t.}\ & &\textrm{Pr}\left(\frac{\mathbf{w}^H_{i}\left(\widehat{\mathbf{h}}_{i}+\mathbf{H}_{i}^{1/2}\mathbf{e}_{i}\right)
\left(\widehat{\mathbf{h}}_{i}+\mathbf{H}_{i}^{1/2}\mathbf{e}_{i}\right)^H\mathbf{w}_{i}}
{\sum_{j =1,j \neq i}^U\mathbf{w}^H_{j}\left(\widehat{\mathbf{h}}_{i}+\mathbf{H}_{i}^{1/2}\mathbf{e}_{i}\right)
\left(\widehat{\mathbf{h}}_{i}+\mathbf{H}_{i}^{1/2}\mathbf{e}_{i}\right)^H
\mathbf{w}_{j}+\sigma^2_i}\geq \gamma_{i}\right) \\
&&& \geq 1-\rho_i, \forall i \in \{ 1, \cdots, U\},
\end{aligned}
\end{equation}
where
$\rho_i\in(0,1]$ is the predefined maximum tolerable probabilities/chances of SINR outages.

Similarly, introducing a new optimization variable $\mathbf{W}_i=\mathbf{w}_i\mathbf{w}_i^H$, using \eqref{sinr_event}, we cast \eqref{chance_ori} as:
\begin{equation}
\begin{aligned}\label{chance_01}
& \displaystyle \min_{\{\mathbf{W}_{i}\}\in \mathbb{H}^{M \times M}} & &
\sum_{i=1}^U \textrm{Tr}\left(\mathbf{W}_{i} \right)\\
& \text{s.\ t.}\ & &\textrm{Pr}\left(f_i(\mathbf{e}_{i})\geq 0\right) \geq 1-\rho_i, \forall i \in \{ 1, \cdots, U\},
\\ & & &
\mathbf{W}_{i}\succeq \mathbf{0},\ \textrm{rank}\left(\mathbf{W}_{i}\right)=1, \ \forall i \in \{ 1, \cdots, U\}.
\end{aligned}
\end{equation}
Although the event $f_i(\mathbf{e}_{i})\geq 0$ is convex, the corresponding probabilistic constraint  in \eqref{chance_01} is neither necessarily convex nor admits a simple closed-form. To overcome the challenge, our goal is to derive convex upper bounds for the chance constraint in \eqref{chance_01}. This convex approximation approach is based on the large deviation inequality, i.e., a Berstein-type inequality, which bounds the probability that a sum of random variables deviates from its mean \cite{Kun-Yu2014}.  To begin with, let us recall the following lemma.
\begin{lemma}[Bernstein-type inequality \cite{Bernstein-type}]\label{bernstein}
Consider the following random variable $f(\mathbf{x})=\mathbf{x}^H\mathbf{Y}\mathbf{x}+2\textrm{Re}\{\mathbf{x}^H\mathbf{u}\}$, where $\mathbf{x}\sim\mathcal{CN}(\mathbf{0},\mathbf{I}_M)$,\footnote{Here, the errors are assumed to be uncorrelated. When the errors are correlated, i.e., the covariance matrix of $\mathbf{x}$ is not an identity matrix, a novel convex approximation approach is needed. This deserves a new research topic. } $\mathbf{Y}\in \mathbb{H}^{M\times M}$, and $\mathbf{u}\in \mathbb{C}^{M\times 1}$. For all $\delta >0$, the following statement holds:
\begin{eqnarray}
\textrm{Pr}\left(f(\mathbf{x})\geq \textrm{Tr}\left( \mathbf{Y}\right)-\sqrt{2\delta}\sqrt{\|\mathbf{Y}\|_F^2+2\|\mathbf{u}\|^2}-\delta s^+(\mathbf{Y})\right) \geq 1 - e^{-\delta}.
\end{eqnarray}
\end{lemma}

With $\delta_i=-\ln{\rho_i}$ and Lemma \ref{bernstein}, the SINR outage constraint $\textrm{Pr}\left( f_i(\mathbf{e}_i)\geq 0\right)\geq 1-\rho_i$ in \eqref{chance_01} can be rewritten as: 
\begin{eqnarray}
\textrm{Tr}\left(\mathbf{H}_{i}^{1/2}\mathbf{\widetilde{B}}_i\mathbf{H}_{i}^{1/2}\right)-\sqrt{2\delta_i}
\sqrt{\|\mathbf{H}_{i}^{1/2}\mathbf{\widetilde{B}}_i\mathbf{H}_{i}^{1/2}\|_F^2+2\|
\mathbf{H}_{i}^{1/2}\mathbf{\widetilde{B}}_i\widehat{\mathbf{h}}_{i}\|^2} \nonumber \\ - \delta_i s^+\left(\mathbf{H}_{i}^{1/2}\mathbf{\widetilde{B}}_i\mathbf{H}_{i}^{1/2}\right)\geq \sigma_i^2-
\widehat{\mathbf{h}}_{i}^H\mathbf{\widetilde{B}}_i\widehat{\mathbf{h}}_{i}. \label{bernstein1}
\end{eqnarray}
Then, by introducing two auxiliary optimization variables $\varrho_i$ and $\vartheta_i$,  and adopting the identity $\sqrt{\|\mathbf{a}\|^2+\|\mathbf{A}\|_F^2}=\|\left[\mathbf{a}^T, \textrm{vec}\left( A\right)^T\right]^T\|$, \eqref{bernstein1} is further recast as the following equivalent constraint:
\begin{eqnarray}
\textrm{Tr}\left(\mathbf{H}_{i}^{1/2}\mathbf{\widetilde{B}}_i\mathbf{H}_{i}^{1/2}\right)
-\sqrt{2\delta_i}\varrho_i-\delta_i \vartheta_i\geq \sigma_i^2-
\widehat{\mathbf{h}}_{i}^H\mathbf{\widetilde{B}}_i\widehat{\mathbf{h}}_{i},\label{const11}\\
\left \Vert\begin{bmatrix}
\sqrt{2}\mathbf{H}_{i}^{1/2}\mathbf{\widetilde{B}}_i\widehat{\mathbf{h}}_{i} \\
\textrm{vec}\left( \mathbf{H}_{i}^{1/2}\mathbf{\widetilde{B}}_i\mathbf{H}_{i}^{1/2}\right)
\end{bmatrix}
\right \Vert\leq \varrho_i,\label{const12} \\
\vartheta_i\mathbf{I}_M+\mathbf{H}_{i}^{1/2}\mathbf{\widetilde{B}}_i\mathbf{H}_{i}^{1/2}  \succeq \mathbf{0},\label{const13}\\
\vartheta_i\geq 0\label{const14}.
\end{eqnarray}

From \eqref{const11}-\eqref{const14}, we can equivalently cast \eqref{chance_01} as:
\begin{equation}
\begin{aligned}\label{chance_02}
& \displaystyle \min_{\{\mathbf{W}_{i}\}\in \mathbb{H}^{M \times M}} & &
\sum_{i=1}^U \textrm{Tr}\left(\mathbf{W}_{i} \right)\\
& \text{s.\ t.}\ & &\eqref{const11},\eqref{const12},\eqref{const13},\eqref{const14}, \ \forall i \in \{ 1, \cdots, U\},
\\ & & &
\mathbf{W}_{i}\succeq \mathbf{0},\ \textrm{rank}\left(\mathbf{W}_{i}\right)=1, \ \forall i \in \{ 1, \cdots, U\}.
\end{aligned}
\end{equation}
Mirroring the discussions in the previous subsections, in the following we will express the constraints of \eqref{chance_02} in the compact form as \eqref{rootSDP}. First, substituting for $\mathbf{\widetilde{B}}_i$ in \eqref{B_i} after some manipulations, one can rewrite \eqref{const11} as:
\begin{eqnarray}
a_i\textrm{Tr}\left(\mathbf{X}_{i,i}\mathbf{W}_i\right)+\sum_{j=1}^U b_j\textrm{Tr}
\left(\mathbf{X}_{i,j}\mathbf{W}_{j}\right)+c_i\geq0,
\end{eqnarray}
    where $a_i=\left( 1+\frac{1}{\gamma_i}\right)$, $b_j=-1$, $\mathbf{X}_{i,i}=\mathbf{X}_{i,j}=\mathbf{H}_i+\widehat{\mathbf{h}}_i\widehat{\mathbf{h}}_i^H$, and $c_i=
-\sqrt{2\delta_i}\varrho_i-\delta_i \vartheta_i - \sigma_i^2$. Next, letting $e_i=\sqrt{2}$, $\mathbf{Z}_i=\mathbf{H}_{i}^{1/2}$, $\mathbf{C}_i=\mathbf{\widetilde{B}}_i$, $\mathbf{z}_i=\mathbf{h}_{i}$, $f_i=\vartheta_i$, $v_i=1$, $\boldsymbol\Lambda_k=\boldsymbol\Psi_k=\mathbf{I}_M$, and $\mathbf{A}_i=\mathbf{I}_M$, using those mappings and relaxing the rank-one constraint on $\mathbf{W}_i$, we recast \eqref{chance_02} as:
\begin{equation}
\begin{aligned}\label{chanced_SDP}
& \displaystyle \min_{\{\mathbf{W}_{i}\}\in \mathbb{H}^{M\times M}} & &
\sum_{i=1}^U \textrm{Tr}\left(\mathbf{A}_{i}\mathbf{W}_{i}\right) \\
& \text{s.\ t.}\ & &a_i\textrm{Tr}\left(\mathbf{X}_{i,i}\mathbf{W}_i\right)+\sum_{j=1}^U b_j\textrm{Tr}
\left(\mathbf{X}_{i,j}\mathbf{W}_{j}\right)+c_i\geq 0, \\
&&& \ \ \ \ \ \  \ \ \ \ \ \ \ \ \ \ \ \ \ \ \ \ \ \ \ \ \ \ \ \ \forall i \in \{1,\cdots,U\},\\
&&& \left \Vert\begin{bmatrix}
e_i\mathbf{Z}_i\mathbf{C}_i
\mathbf{z}_i \\
\textrm{vec}\left( \mathbf{Z}_i\mathbf{C}_i\mathbf{Z}_i\right)
\end{bmatrix}
\right \Vert\leq \varrho_i,\ \forall i  \in \{1,\cdots,U\},\\
&&&  \ f_i\mathbf{I}_M+v_i\mathbf{D}_i\sum_{k=1}^N\boldsymbol\Lambda_k\mathbf{E}_i\boldsymbol\Psi_k\widetilde{\mathbf{D}}_i\succeq \mathbf{0},\ f_i \geq 0, \\
&&& \ \ \ \ \ \  \ \ \ \ \ \ \ \ \ \ \ \ \ \ \ \ \ \ \ \ \ \ \ \forall i  \in \{1,\cdots,U\},\\
&&& \mathbf{W}_{i} \succeq \mathbf{0}, \ \forall i \in \{1,\cdots,U\}.
\end{aligned}
\end{equation}

Using the Schur complement on the second constraint with some mathematical manipulations, one can rewrite \eqref{chanced_SDP} as:
\begin{equation}
\begin{aligned}\label{chanced_LMI}
& \displaystyle \min_{\{\mathbf{W}_{i}\}\in \mathbb{H}^{M\times M}} & &
\sum_{i=1}^U \textrm{Tr}\left(\mathbf{A}_{i}\mathbf{W}_{i}\right) \\
& \text{s.\ t.}\ & &a_i\textrm{Tr}\left(\mathbf{X}_{i,i}\mathbf{W}_i\right)+\sum_{j=1}^U b_j\textrm{Tr}
\left(\mathbf{X}_{i,j}\mathbf{W}_{j}\right)+c_i\geq 0,\\
&&& \ \ \ \ \ \ \ \ \ \ \ \ \ \ \ \ \ \ \ \ \ \ \ \ \ \ \ \ \ \forall i \in \{1,\cdots,U\},\\
&&& \begin{bmatrix}\varrho_i \mathbf{I}_{M^2+M} & \begin{bmatrix}
e_i\mathbf{Z}_i\mathbf{C}_i
\mathbf{z}_i \\
\textrm{vec}\left( \mathbf{Z}_i\mathbf{C}_i\mathbf{Z}_i\right)
\end{bmatrix}\\
\begin{bmatrix}
e_i\mathbf{Z}_i\mathbf{C}_i
\mathbf{z}_i \\
\textrm{vec}\left( \mathbf{Z}_i\mathbf{C}_i\mathbf{Z}_i\right)
\end{bmatrix}^H&\varrho_i\end{bmatrix}
\succeq \mathbf{0}, \\
&&& \ \ \ \ \ \ \ \ \ \ \ \ \ \ \ \ \ \ \ \ \ \ \ \ \ \ \   \forall i  \in \{1,\cdots,U\},\\
&&&  \ f_i\mathbf{I}_M+v_i\mathbf{D}_i\sum_{k=1}^N\boldsymbol\Lambda_k\mathbf{E}_i\boldsymbol\Psi_k\widetilde{\mathbf{D}}_i\succeq \mathbf{0}, \ \forall i  \in \{1,\cdots,U\},\\
&&& \mathbf{W}_{i} \succeq \mathbf{0}, \ \forall i \in \{1,\cdots,U\}.
\end{aligned}
\end{equation}
It can be observed that problem \eqref{chanced_LMI} is a sub-class optimization problem of the general framework \eqref{rootSDP}, i.e., including constraints $C_1$, $C_4(a)$, $C_5$, and $C_6$ with $L_1=L_4(a)=L_5=U$. Therefore, Corollary~\ref{cor01} implies that \eqref{chanced_LMI} yields rank-one optimal solution when it is feasible and its optimal solution is not trivial. Hence, the optimal solution of \eqref{chanced_LMI} is also the optimal solution of \eqref{chance_02}. Consequently, the SDR \eqref{chanced_LMI} is equivalent to the original QCQP \eqref{chance_02}.

Keeping $C_1$, $C_4(a)$, $C_5$, and $C_6$ with $L_1=L_4(a)=L_5=U$ in the general framework \eqref{rootSDP}, one can derive the complexity of \eqref{chanced_LMI} from Lemma~\ref{lemSDPcog} as follows.

\begin{cor}
    The computational complexity to attain $\varepsilon$-solution to \eqref{chanced_LMI}  is on the order of:
\begin{align}
    \ln{\left(\varepsilon^{-1}\right)}\sqrt{\beta_3\Big(\mathcal{M}\Big)}&\Big[ C_{\text{form},3}+C_{\text{fact},3}\Big],
\end{align}
where $\beta_3(\mathcal{M})=2UM+U(M^2+M+2)$, $C_{\text{form},3}=M^2\Big[ U+2UM^3+U\Big( M^2+M+1\Big)^3\Big]+M^4\Big[U+2UM^2+U\Big( M^2+M+1\Big)^2\Big]$, and $C_{\text{fact},3}=4M^6$.
\end{cor}
\subsection{Reconfigurable-Intelligent-Surface-Aided Beamforming}
Consider a communication system consisting of an $M$-antenna BS serving $U$ single antenna mobile users in which the direct communication link between the BS and its mobile users is blocked, e.g., due to high building etc., \cite{Tuan2021}. To circumvent the problem, an $N$-reflective-elements reconfigurable intelligent surface (RIS) is utilized to support the communication. Let $\mathbf{H} = [\mathbf{h}_1, \ldots, \mathbf{h}_N] \in \mathbb{C}^{M \times N}$ and $\mathbf{g}_i =[g_{i1}, \ldots, g_{iN}]^T \in \mathbb{C}^{N \times 1}$ denote the channel coefficients between the BS and the RIS and those between the RIS and the $i$-th user, respectively.

Let $x_i$, i.e., $\mathbb{E}[|x_i|^2]=1$, and $\mathbf{w}_i \in \mathbb{C}^{M\times 1}$, respectively, represent the data symbol and the beamforming vector for the $i$-th user. Each reflective element  of the RIS generates a phase shift to support the communication between the BS and the mobile users. Let $\theta_n$ be the phase shift at the $n$-th reflective element and let $\pmb{\theta} =[\theta_1, \ \theta_2, \ \cdots, \ \theta_N]^T$ denote the phase-shift coefficients generated by the IRS with $|\theta_n| \leq 1, \forall n = 1, \ldots, N$. The signal received by user~$i$ can be written as:
\begin{eqnarray}\label{re01}
y_i&=&\mathbf{g}_i^H \textrm{diag}(\pmb{\theta})^H \mathbf{H}^H \mathbf{w}_{i} x_{i} +\mathbf{g}_i^H \textrm{diag}(\pmb{\theta})^H \mathbf{H}^H \sum_{j=1,j \neq i}^{U}\mathbf{w}_{j} x_{j}+ n_i,\nonumber \\
&=&\pmb{\theta}^H \mathbf{G}_i^H \mathbf{w}_{i} x_{i} +\pmb{\theta}^H \mathbf{G}_i^H \sum_{j=1,j\neq i}^{U}\mathbf{w}_{j} x_{j} + n_i,
\end{eqnarray}
where $\mathbf{G}_i^H=\textrm{diag}(\mathbf{g}_i^{\ast})\mathbf{H}^H \in \mathbb{C}^{N \times M}$ and $n_i\sim \mathcal{CN}(0,\sigma^2)$ is the additive noise at the $i$-th user. Let $\{\mathbf{w}_{i}\}=\{\mathbf{w}_1, \mathbf{w}_2,\cdots, \mathbf{w}_U\}$ be the set of beamforming vectors, the SINR at user~$i$, denoted by $\textrm{SINR}_i( \{\mathbf{w}_{i}\},\pmb{\theta})$, can be stated as:
\begin{equation}\label{sinr01}
\textrm{SINR}_i\left( \{\mathbf{w}_{i}\},\pmb{\theta}\right) =\frac{|\pmb{\theta}^H \mathbf{G}_i^H \mathbf{w}_i|^2}{ \sum\limits_{j=1,j\neq i}^{U}|\pmb{\theta}^H \mathbf{G}_i\mathbf{w}_{j}|^2+\sigma_i^2}.
\end{equation}
The optimization is formulated as:
\begin{equation} \label{IRS_PerfectCSI}
\begin{aligned}
& \underset{ \{\mathbf{w}_{i} \},\ \pmb{\theta} }{\textrm{min}} & &
\sum_{i=1}^U   \mathbf{w}_{i}^H\mathbf{w}_i\\
& \mbox{s.\ t.}\ & & \textrm{SINR}_{i}\left( \{\mathbf{w}_{i}\},\pmb{\theta}\right) \geq \gamma_i, \forall i, \\
&&& |\theta_n| \leq 1, \forall n. 
\end{aligned}
\end{equation}
Let $\mathbf{W}_i=\mathbf{w}_i\mathbf{w}_i^H$ and $\boldsymbol\Theta=\pmb{\theta}\pmb{\theta}^H$, after some manipulations, one can rewrite \eqref{IRS_PerfectCSI} as:
\begin{equation}
\begin{aligned}\label{prob_IRS}
& \underset{ \{\mathbf{W}_{i} \},\ \mathbf{\Theta}}{\textrm{min}}  & & \textrm{tr}\left(\sum_{i=1}^U \mathbf{W}_{i}\right)\\
& \mbox{s. \ t.}\ & &\left(1+\frac{1}{\gamma_i} \right) \textrm{Tr}\left(\mathbf{G}_{i}\boldsymbol\Theta\mathbf{G}_i^H\mathbf{W}_{i}\right)-\sum_{j=1}^U\textrm{Tr}
\left(\mathbf{G}_{i}\boldsymbol\Theta\mathbf{G}_i^H\mathbf{W}_{j}\right)\\
&&&-\sigma_i^2\geq0,\ \forall i  \in \{1,\cdots,U\},\\ 
&&& \mathbf{W}_i  \succeq \mathbf{0}, \ \mathrm{rank}(\mathbf{W}_i) = 1,\ \forall i  \in \{1,\cdots,U\},\\
&&& \textrm{diag}\left(\textrm{diag}\left(\mathbf{\Theta}\right)\right)   \preceq \mathbf{I}_N, \  \mathbf{\Theta} \succeq \mathbf{0}, \ \mathrm{rank}(\mathbf{\Theta}) = 1.
\end{aligned}
\end{equation}
The problem is non-convex with respect to $\mathbf{W}_i$ and $\boldsymbol\Theta$ due to the fact that the first constraint affinelly depends on of both  $\mathbf{W}_i$ and $\boldsymbol\Theta$. As  $\mathbf{W}_i$ and $\boldsymbol\Theta$ are two independent variables, we adopt an alternating optimization approach \cite{Tuan2021} to solve \eqref{prob_IRS}. To that end, starting from an initial value of the reflecting coefficients $\boldsymbol\Theta^{(0)}$, the following sub-problem will be solved at the $k$-th iteration:
\begin{equation}
\begin{aligned}\label{prob_IRS_W}
& \underset{ \{\mathbf{W}_{i} \},}{\textrm{min}}  & & \textrm{Tr}\left(\sum_{i=1}^U \mathbf{W}_{i}\right)\\
& \mbox{s. \ t.}\ & &\left(1+\frac{1}{\gamma_i} \right) \textrm{Tr}\left(\mathbf{G}_{i}\boldsymbol\Theta^{(k-1)}\mathbf{G}_i^H\mathbf{W}_{i}\right)\\
&&&-\sum_{j=1}^U\textrm{Tr}
\left(\mathbf{G}_{i}\boldsymbol\Theta^{(k-1)}\mathbf{G}_i^H\mathbf{W}_{j}\right)-\sigma_i^2\geq0, \forall i  \in \{1,\cdots,U\}\\ 
&&& \mathbf{W}_i  \succeq \mathbf{0}, \ \mathrm{rank}(\mathbf{W}_i) = 1, \forall i  \in \{1,\cdots,U\}.
\end{aligned}
\end{equation}
In the following we show that the SINR constraint in \eqref{prob_IRS_W} can be expressed the form of $C1$ in  \eqref{rootSDP}. To that end, we map the notations used in \eqref{prob_IRS_W} to those used in \eqref{rootSDP} as follows. First, we denote $a_i=\left(1+\frac{1}{\gamma_i} \right)$, $b_j=-1$, $\mathbf{X}_{i,i}=\mathbf{X}_{i,j}=\mathbf{G}_{i}\boldsymbol\Theta^{(k-1)}\mathbf{G}_i^H$, and $c_i=-\sigma_i^2$.  Relaxing the rank-one constraint on $\mathbf{W}_i$, utilizing those mappings, and letting $\mathbf{A}_i=\mathbf{I}_M$, \eqref{prob_IRS_W} can be rewritten as:
\begin{equation}
\begin{aligned}\label{prob_IRS_W_relax}
& \underset{ \{\mathbf{W}_{i} \},}{\textrm{min}}  & & \textrm{Tr}\left(\sum_{i=1}^U \mathbf{A}_i\mathbf{W}_{i}\right)\\
& \mbox{s. \ t.}\ & &a_i\textrm{Tr}\left(\mathbf{X}_{i,i}\mathbf{W}_i\right)+\sum_{j=1}^U b_j\textrm{Tr}
\left(\mathbf{X}_{i,j}\mathbf{W}_{j}\right)+c_i\geq0,\\
&&& \ \ \ \ \ \ \ \ \ \ \ \ \ \ \ \ \ \ \ \ \ \ \ \ \ \ \ \ \ \ \ \forall i  \in \{1,\cdots,U\} \\ 
&&& \mathbf{W}_i  \succeq \mathbf{0}, \ \forall i  \in \{1,\cdots,U\}.
\end{aligned}
\end{equation}
It can be seen that \eqref{prob_IRS_W_relax} is a sub-class optimization problem of the general framework \eqref{rootSDP}. Hence, if the conditions stated in Corollary~\ref{cor01} are met, \eqref{prob_IRS_W_relax}  yields  rank-one optimal solution which is the same as that of \eqref{prob_IRS_W}.\footnote{In a power-domain NOMA RIS system, where superposition coding and successive interference cancellation are, respectively, implemented at the transmitter and receiver, the corresponding power optimization problem, e.g., \cite{Li2022}, can also be written in the form of \eqref{prob_IRS_W_relax}. Hence, the result from Corollary~\ref{cor01} still holds true.} Therefore, the SDR \eqref{prob_IRS_W_relax} is equivalent to the original QCQP \eqref{prob_IRS_W}.

The reflecting coefficients $\boldsymbol\Theta^{(k)}$ is then updated from the optimal solution of \eqref{prob_IRS_W_relax} at $k$-th iteration, i.e., $\{\mathbf{W}_i^{(k)}\}$, by solving the following subproblem:
\begin{equation}
\begin{aligned}\label{prob_IRS_Theta}
& \underset{ }{\textrm{find}}  & & \mathbf{\Theta}\\
& \mbox{s. \ t.}\ & &\left(1+\frac{1}{\gamma_i} \right) \textrm{Tr}\left(\boldsymbol\Theta\mathbf{G}_i^H\mathbf{W}_{i}^{(k)}\mathbf{G}_{i}\right)-\sum_{j=1}^U\textrm{Tr}
\left(\boldsymbol\Theta\mathbf{G}_i^H\mathbf{W}_{j}^{(k)}\mathbf{G}_{i}\right)\\&&&-\sigma_i^2\geq0,\ \forall i  \in \{1,\cdots,U\},\\ 
&&& \textrm{diag}\left(\textrm{diag}\left(\mathbf{\Theta}\right)\right)   \preceq \mathbf{I}_N, \  \mathbf{\Theta} \succeq \mathbf{0}, \ \mathrm{rank}(\mathbf{\Theta}) = 1.
\end{aligned}
\end{equation}
We introduce the following problem to find optimal solution for problem \eqref{prob_IRS_Theta} \cite{Tuan2021}:
\begin{equation}
\begin{aligned}\label{prob_IRS_Theta_relax}
& \underset{ \mathbf{\Theta}}{\textrm{min}}  & & \textrm{Tr}\left( \boldsymbol\Theta\right)\\
& \mbox{s. \ t.}\ & &\left(1+\frac{1}{\gamma_i} \right) \textrm{Tr}\left(\boldsymbol\Theta\mathbf{G}_i^H\mathbf{W}_{i}^{(k)}\mathbf{G}_{i}\right)-\sum_{j=1}^U\textrm{Tr}
\left(\boldsymbol\Theta\mathbf{G}_i^H\mathbf{W}_{j}^{(k)}\mathbf{G}_{i}\right)\\&&&-\sigma_i^2\geq0,\ \forall i  \in \{1,\cdots,U\},\\ 
&&& \textrm{diag}\left(\textrm{diag}\left(\mathbf{\Theta}\right)\right)   \preceq \mathbf{I}_N, \\
&&&\mathbf{\Theta} \succeq \mathbf{0}.
\end{aligned}
\end{equation}
Similarly, we can rewrite the SINR constraint in \eqref{prob_IRS_Theta_relax} as $C1$ with $a_i=\left(1+\frac{1}{\gamma_i} \right)$, $b_j=-1$, $\mathbf{X}_{i,i}=\mathbf{G}_{i}^H\mathbf{W}_i^{(k)}\mathbf{G}_i$, $\mathbf{X}_{i,j}=\mathbf{G}_{i}^H\mathbf{W}_j^{(k)}\mathbf{G}_i$, and $c_i=-\sigma_i^2$. In the sequence, we express the phase amplitude constraint, i.e., the second constraint, in \eqref{prob_IRS_Theta_relax} as $C5$. One can write
\begin{eqnarray}
\textrm{diag}\left(\textrm{diag}\left(\mathbf{\Theta}\right)\right)  =\sum_{k=1}^N\boldsymbol\Omega_k \mathbf{\Theta} \boldsymbol\Omega_k,
\end{eqnarray}
where $\boldsymbol\Omega_k$ is an $N\times N$ matrix contains all zeros but $1$ at the $(k,k)$-th entry. Therefore, the phase amplitude constraint can be rewrite as:
\begin{eqnarray}
\mathbf{I}_N-\sum_{k=1}^N\boldsymbol\Omega_k \mathbf{\Theta} \boldsymbol\Omega_k \succeq \mathbf{0}.
\end{eqnarray}
Let $f=f_i=1$, $v=v_i=-1$, $\mathbf{D}=\mathbf{D}_i=\mathbf{I}_N$, $\widetilde{\mathbf{D}}=\widetilde{\mathbf{D}}_i=\mathbf{I}_N$, $\boldsymbol\Psi_k=\boldsymbol\Lambda_k=\boldsymbol\Omega_k$, and $\mathbf{E}_i=\boldsymbol\Theta$, one can write the phase amplitude constraint of \eqref{prob_IRS_Theta_relax} as $C5$. Finally, let $\mathbf{A}_i=\mathbf{A}=\mathbf{I}_N$, we rewrite \eqref{prob_IRS_Theta_relax} as:
\begin{equation}
\begin{aligned}\label{prob_IRS_Theta_relax_std}
& \underset{ \mathbf{\Theta}}{\textrm{min}}  & & \textrm{Tr}\left(\mathbf{A}\mathbf{\Theta}\right)\\
& \mbox{s. \ t.}\ & &a_i \textrm{Tr}\left(\mathbf{X}_{i,i}\boldsymbol\Theta\right)+\sum_{j=1}^Ub_i\textrm{Tr}
\left(\mathbf{X}_{i,j}\boldsymbol\Theta\right)+c_i\geq0,\ \forall i,\\ 
&&& f\mathbf{I}_N+v\mathbf{D}\sum_{k=1}^N\boldsymbol\Lambda_k\mathbf{E}_i\boldsymbol\Psi_k\widetilde{\mathbf{D}}\succeq \mathbf{0}, \\
&&&\mathbf{\Theta} \succeq \mathbf{0}.
\end{aligned}
\end{equation}

It is clear that \eqref{prob_IRS_Theta_relax_std} is a sub-class optimization problem of the general framework \eqref{rootSDP}, i.e., including $C_1$, $C_5$, and $C_6$. Hence, according to Corollary~\ref{cor01}, \eqref{prob_IRS_Theta_relax_std}  yields  rank-one optimal solution if it is feasible and its optimal solution is not trivial. Therefore, \eqref{prob_IRS_Theta_relax_std} is an approximation problem of \eqref{prob_IRS_Theta} as every feasible solution of \eqref{prob_IRS_Theta_relax_std} is also feasible for \eqref{prob_IRS_Theta} \cite{TuanTGCN2017,Tuan2021}.

Let $\mathcal{K}$ be the number of iterations between solving two sub-problems \eqref{prob_IRS_W_relax} and \eqref{prob_IRS_Theta_relax_std}, $\mathcal{T}_{\mathbf{W}}$ be the complexity of \eqref{prob_IRS_W_relax}, and $\mathcal{T}_{\mathbf{\Theta}}$ be the complexity of \eqref{prob_IRS_Theta_relax_std}. The complexity of the alternating optimization approach is given as follows.

\begin{cor}
    The computational complexity to attain $\varepsilon$-solution to \eqref{prob_IRS_W_relax} and \eqref{prob_IRS_Theta_relax_std}  is on the order of:
    \begin{equation}
    \mathcal{K}\Big( \mathcal{T}_{\mathbf{W}}+\mathcal{T}_{\mathbf{\Theta}}\Big),
    \end{equation}
where $\mathcal{T}_{\mathbf{W}}$ is calculated as in \eqref{complex01},
\begin{align}
        \mathcal{T}_{\mathbf{\Theta}}&=\ln{\left(\varepsilon^{-1}\right)}\sqrt{\beta_4\Big(\mathcal{M}\Big)}\Big[ C_{\text{form},4}+C_{\text{fact},4}\Big],\label{ttheta}\\
    \beta_4(\mathcal{M})&=U+2M,\\
    C_{\text{form},4}&=M^2\Big[ 1+2M^3\Big]+M^4\Big[1+2M^2\Big],\\
    C_{\text{fact},4}&=3M^6.
\end{align}
\end{cor}

 \begin{proof}
 Since two sub-problems \eqref{prob_IRS_W_relax} and \eqref{prob_IRS_Theta_relax_std} are iteratively solved in $\mathcal{K}$ iterations, the complexity is $\mathcal{K}\Big( \mathcal{T}_{\mathbf{W}}+\mathcal{T}_{\mathbf{\Theta}}\Big)$. Comparing \eqref{cog_perfect_2} and \eqref{prob_IRS_W_relax}, one can conclude that the two problems have the same structure. Therefore, \eqref{cog_perfect_2} and \eqref{prob_IRS_W_relax} have the same complexity. Hence, $\mathcal{T}_{\mathbf{W}}$ can be calculated as \eqref{complex01}. Considering the general framework \eqref{rootSDP} with $U$ constraints $C_2$, one constraint $C_5$, and one constraint $C_6$, using Lemma~\ref{lemSDPcog},  the complexity of  \eqref{prob_IRS_Theta_relax_std} $\mathcal{T}_{\mathbf{\Theta}}$ can be derived as in \eqref{ttheta}.
\end{proof}
\section{Numerical Results}
In this section, we show some numerical results to confirm the findings of Theorem 1 and Corollary 1, i.e., to verify the optimal rank-one solutions of the proposed framework. Particularly, we test the tightness, i.e., yielding rank-one optimal solutions, of the transmit beamforming with perfect CSI as in problem \eqref{cog_perfect_2}, hereafter referred to as Perfect CSI approach, the transmit beamforming with imperfect CSI as in \eqref{imperfectCIS_SDP}, hereafter referred to as S-procedure approach, the chance-constraint problem \eqref{chanced_LMI}, hereafter referred to as Chance approach, and the RIS-Aided beamforming approach as in problems \eqref{prob_IRS_W_relax} and \eqref{prob_IRS_Theta_relax_std}, hereafter referred to as RIS-Aided approach.
\subsection{Evaluation Metrics and Setup}
In order to evaluate the tightness of the SDR approach, we adopt the rank one test (ROT) ratio for the beamforming matrix defined as  \cite{Song_lou}:
 \begin{equation}
\text{ROT}_W=
\displaystyle \max_{i} \frac{\sum_{k=2}^{U}\lambda_k(\mathbf{W}_i)}{\lambda_1(\mathbf{W}_i)},\label{ROT}
\end{equation}
where $\lambda_k(\mathbf{W}_i)$ is the $k$-th largest eigenvalue of $\mathbf{W}_i$. $\mathbf{W}_i$ is a rank-one matrix if its ROT is close to zero. Similarly the ROT for the phase-shift coefficient matrix as:
\begin{equation}
\text{ROT}_{\boldsymbol\Theta}=
 \frac{\sum_{k=2}^{N}\lambda_k(\boldsymbol\Theta)}{\lambda_1(\boldsymbol\Theta)}.\label{ROTP}
\end{equation}

In our evaluations, the BS serves two users, i.e., $U=2$. The evaluations are carried out with different numbers of BS's antennas and RIS's reflective elements. The coefficients of the channels are modeled as  $\mathbf{h}_{i}\sim\mathcal{CN}(\mathbf{0},\boldsymbol{\Delta})$ where the diagonal elements of the covariance matrix $\boldsymbol{\Delta}_{i,i}=1,\ \forall i\in \{1,\cdots,M\}$. The proposed framework is investigated under  uncorrelated and correlated channel models. In an uncorrelated channel model, it is assumed that the off diagonal elements of the covariance matrix $\boldsymbol{\Delta}_{i,j}=0$, $\forall i,k\in \{1,\cdots,M\}$ and $i \neq j$, i.e., $\mathbf{Y}=\mathbf{I}_M$, \cite{Kun-Yu2014,Song_lou}. Whereas, in a correlated channel model, it is assumed that $\boldsymbol{\Delta}_{i,j}\neq0$, $\forall i,k\in \{1,\cdots,M\}$ and $i \neq j$. Without loss of generality, the off diagonal elements are set to be either $0.4$ or $0.6$ for the correlated channel model. The channel estimation error covariance matrix $\mathbf{H}_{i}$ is modeled as $\epsilon^2 \mathbf{I}_M$ where $\epsilon^2$ is set to be $0.002$. The noise variances of all users are set to be $0.001$. The SINR outage probability is 0.1, i.e., $\rho_i =\rho=0.1 \ \forall i \in\{1,\cdots,U\}$. In order to provide a fair comparison between the S-procedure and the Chance approaches, the channel uncertainty bound in the S-procedure approach, i.e., $r$, is set as $r=\sqrt{\phi_{m}^{-1}\frac{\left( 1-\rho\right)}{2}}$ where $\phi_{m}^{-1}\left(\cdot\right)$ is the inverse cumulative distribution function of a Chi-square random variable with $m = 2M$ degrees of freedom \cite{Kun-Yu2014}. 
\begin{figure}[t]
		\centering
		\includegraphics[width=.45\textwidth]{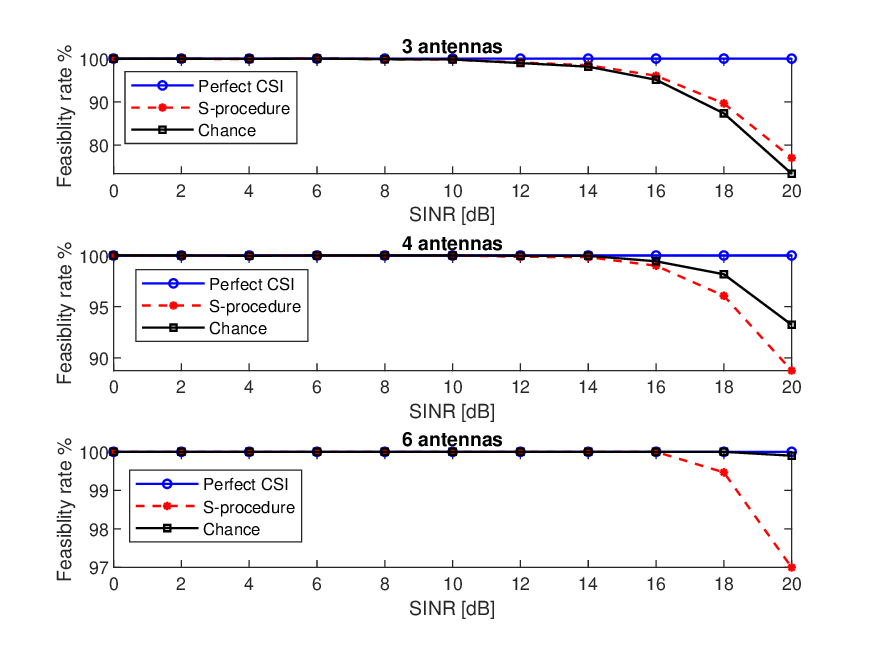} 
		\vspace*{-0.2cm}
		\caption{Feasibility rate versus user target SINR with different number of BS's antennas. Uncorrelated channel model.}
		\label{Fea}
		\vspace*{-0.3cm}
	\end{figure}
	
\subsection{Evaluation for Perfect CSI Approach, S-procedure Approach, and Chance Approach}
\subsubsection{Uncorrelated Channel Model}
In this experiment, Monte Carlo simulations are performed with $\numprint{3000}$ independently channel realizations for each SINR target ranging from $0$ dB to $20$ dB with a step of 2.

Fig.~\ref{Fea} shows the feasibility rate, i.e., percentage of feasible channel realizations, versus the user target SINRs of the under investigated approaches. Simulation results indicate that the Perfect CSI approach is feasible for all $\numprint{3000}$ channel realizations with all antenna setups of $M=3$, $M=4$, and $M=6$ over the observed SINR range. On the other hand, depending on the system setup, e.g., the number of users, the number of antenna elements or required SINR, the robust beamforming approaches, i.e., the S-procedure and Chance approaches, may have  infeasible channel realizations due to imperfect CSI. For example, at the required SINR of 20 dB, the feasibility's channel realization percentages of the S-procedure and Chance approaches are, respectively, $77 \ \%$ and $73.4\ \%$ with 3 antenna elements while those are, respectively, $88.8 \ \%$ and $93.2 \ \%$ with 4 antenna elements. Interestingly, when the number of antenna elements increases to 6, the number of feasible channel realizations of the Chance approach closely follows that of the Perfect CSI approach. This is due to the following facts. First, the derivation of the SDR for the Chance approach does not introduce any extra constraint to the original QCQP. Second,
increasing the number of antenna elements enlarges the feasibility region of the Chance approach hence allows it to closely reach that of the Perfect CSI approach.
	\begin{figure}[t]
		\centering
		\includegraphics[width=.45\textwidth]{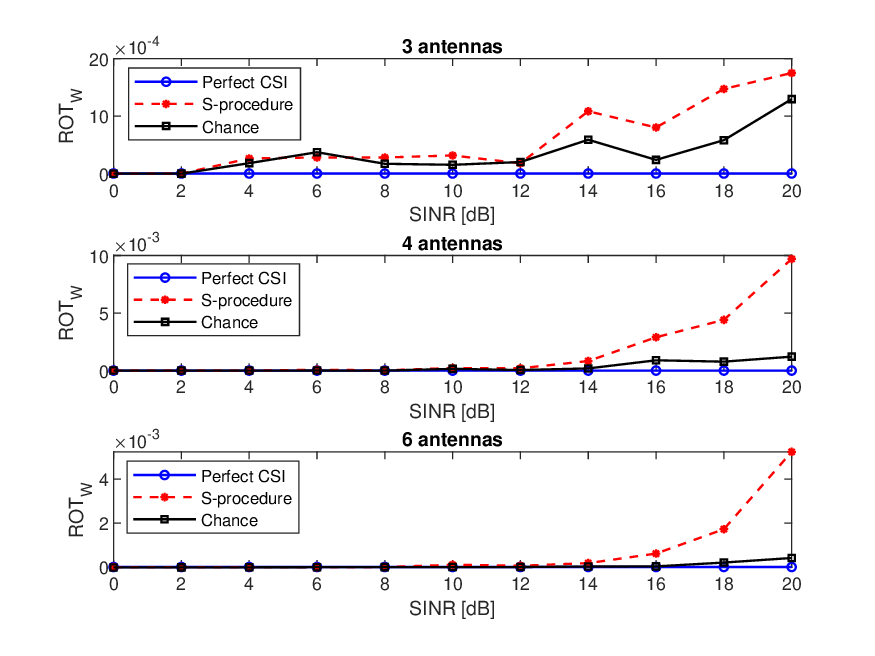} 
		\vspace*{-0.2cm}
		\caption{ROT versus user target SINR with different number of BS's antennas. Uncorrelated channel model.}
		\label{RoTfig}
		\vspace*{-0.3cm}
	\end{figure}

Fig.~\ref{RoTfig} illustrates the ROTs  of the beamforming matrices of the three approaches with different number of antennas versus the user target SINRs. Those ROTs are obtained when the corresponding optimization problems of the three approaches are feasible. It is clear from Fig.~\ref{RoTfig} that all ROTs are close to zero which imply that the optimal solutions to problems \eqref{cog_perfect_2}, \eqref{imperfectCIS_SDP} and \eqref{chanced_LMI} are all rank-one matrices. This confirms Theorem 1 and Corollary 1. Observing the SINR range beyond $12$ dB in Fig.~\ref{RoTfig} reveals a fact that a problem associated with $C1$ and $C6$ constraints has the lowest ROTs followed by a problem associated with $C3$ and $C6$ constraints while a problem associated with $C4(a)$, $C5$, and $C6$ constraints has the highest ROTs. A similar performance trend in terms of the feasibility rate can aslo been seen from Fig.~\ref{Fea} where a problem with $C3$ and $C6$ constraints has the highest feasibility rate followed by a problem with $C3$ and $C6$ constraints and a problem with $C4(a)$, $C5$, and $C6$ constraints. The results in Figs.~\ref{Fea} and \ref{RoTfig} again confirm a fact in the literature that a beamforming approach has the best performance, i.e., regarding the average transmit power consumption and the feasibility rate, when perfect CSI is available, see e.g., \cite{Kun-Yu2014}. In the presence of imperfect CSI, its performance degrades and the performance of the Chance approach outperforms that of the S-procedure approach, see e.g., \cite{Kun-Yu2014}.
\begin{figure}[t]
		\centering
		\includegraphics[width=.45\textwidth]{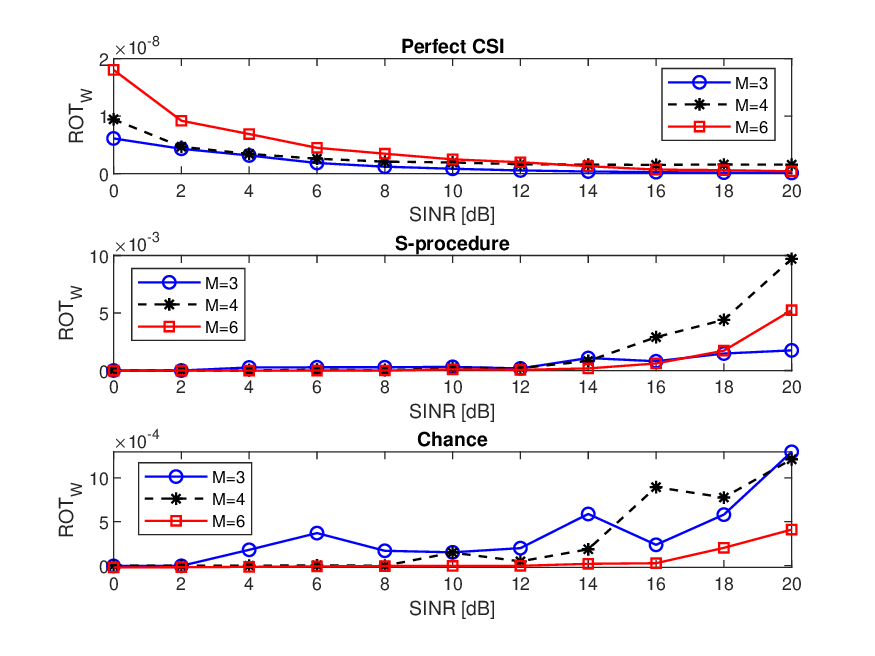} 
		\vspace*{-0.2cm}
		\caption{ROT versus user target SINR with different number of BS's antennas. Uncorrelated channel model.}
		\label{RoTfig2}
		\vspace*{-0.3cm}
	\end{figure}
	
 Fig.~\ref{RoTfig2} brings a different view of Fig.~\ref{RoTfig} where the ROTs are shown for each beamforming approach. It can be observed from Fig.~\ref{RoTfig2} that there are slightly differences in the ROT with different setups of $M$ for the observed SINR range. This reveals the fact that the number of antennas and the SINR level do not have significant impact on the ROTs.
 \subsubsection{Correlated Channel Model}In this experiment, Monte Carlo simulations are performed with $\numprint{500}$ independently channel realizations for each SINR target ranging from $0$ dB to $20$ dB with a step of 2.
 \begin{figure}[t]
		\centering
	\includegraphics[width=.45\textwidth]{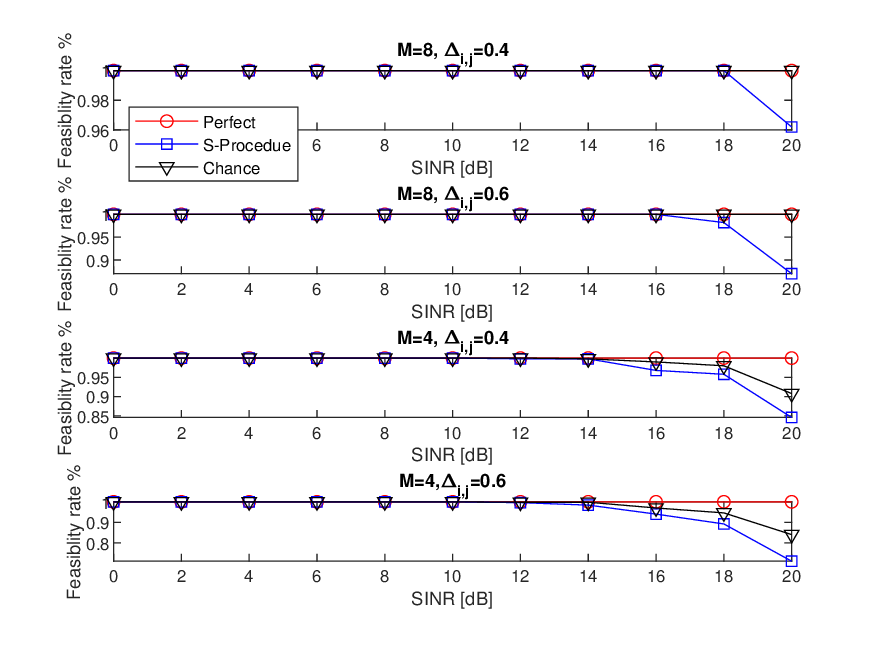} 
		\vspace*{-0.2cm}
		\caption{Feasibility rate versus user target SINR with different number of BS's antennas and values of $\boldsymbol{\Delta}_{i,j}$. Correlated channel model.}
		\label{Corralated_Fe}
		\vspace*{-0.3cm}
	\end{figure}
 
Fig.~\ref{Corralated_Fe} illustrates the feasibility rate versus the user target SINRs of the three transmit beamforming approaches with different number of BS's antennas and values of $\boldsymbol{\Delta}_{i,j}$. Similar behaviours the three beamforming approaches' feasibility rates, as in the uncorrelated channel model, are observed here. The Perfect CSI approach is feasible for all channel realizations under the observed setup. The S-procedure and Chance approaches are feasible in the SINR range from 0 dB to 14 dB. From 16 dB to 20 dB, a higher channel correlated value $\boldsymbol{\Delta}_{i,j}$ results in a lower feasibility rate. For example, the feasibility rate of the Chance approach decreases from 84.6 \% to 71 \% when $\boldsymbol{\Delta}_{i,j}$ increases from 0.4 to 0.6 with $M=4$.
 \begin{figure}[t]
		\centering
		\includegraphics[width=.45\textwidth]{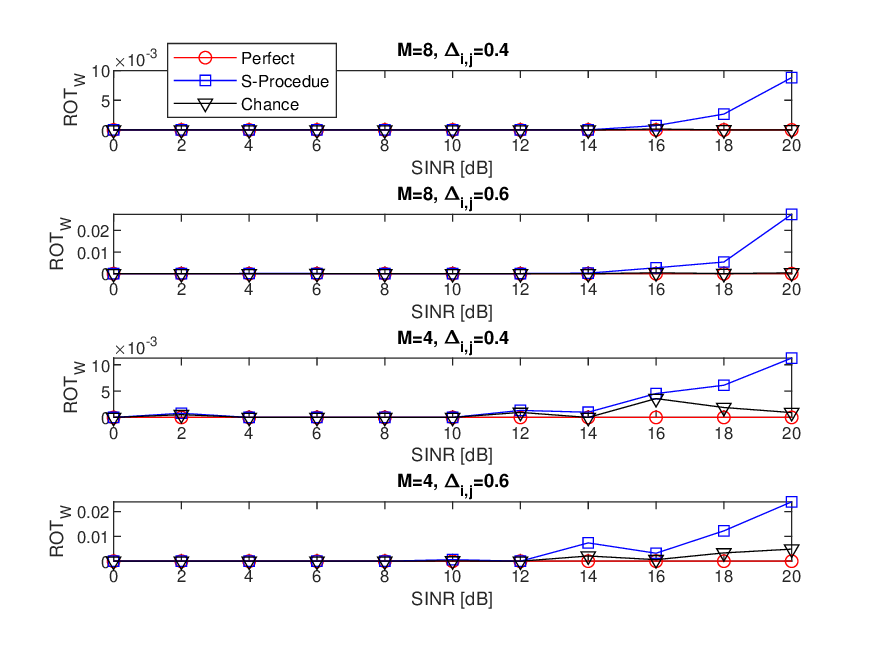} 
		\vspace*{-0.2cm}
		\caption{ROT versus user target SINR with different number of BS's antennas and values of $\boldsymbol{\Delta}_{i,j}$. Correlated channel model.}
		\label{Corralated_ROT}
		\vspace*{-0.3cm}
	\end{figure}

 Fig.~\ref{Corralated_ROT} shows the ROTs of the three beamforming approaches versus the user target SINRs with different numbers of BS's antennas and values of $\boldsymbol{\Delta}_{i,j}$. It can be seen that all ROTs are close to zero. For example, the ROTs of the worst-performance approach, i.e., the S-Procedure approach, are in the order of $10^{-2}$ and $10^{-3}$, respectively, for $\boldsymbol{\Delta}_{i,j}=0.6$ and $\boldsymbol{\Delta}_{i,j}=0.4$. The results confirm Theorem 1 and Corollary 1. The results on Fig.~\ref{Corralated_ROT} also indicate that $C3$ constraints, which associate with the S-Procedure approach, are more sensitive with the values of $\boldsymbol{\Delta}_{i,j}$ than the other constraints like $C4a$ and $C5$, which associate with the Perfect CSI and Chance approaches. The other observations of these schemes in the uncorrelated channel model, i.e., the discussions of Fig.~\ref{RoTfig}, are still valid for the correlated channel model.
\subsection{Evaluation for RIS-Aided Approach}
\subsubsection{Uncorrelated Channel Model}
In this experiment, the RIS-Aided approach is investigated. Monte Carlo simulations are performed with $\numprint{1000}$ independently channel realizations for each SINR target ranging from $0$ dB to $20$ dB with a step of 2. The simulation results indicate that all channel realizations are feasible for the observed SINR range. 
		\begin{figure}[t]
		\centering
		\includegraphics[width=.45\textwidth]{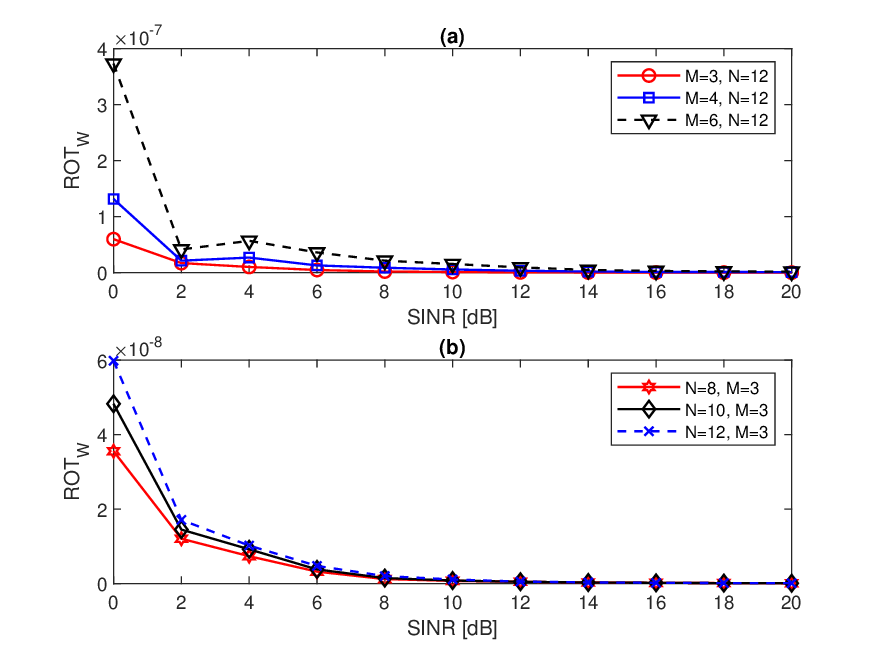} 
		\vspace*{-0.2cm}
		\caption{ROT of the beamforming matrix versus user target SINR. (a): with different numbers of BS's antennas and a fixed number of reflective elements $N=12$. (b):  with different numbers of reflective elements and a fixed number of BS's antenna $M=3$. Uncorrelated channel model.}
		\label{ROTWIRS}
		\vspace*{-0.3cm}
	\end{figure}
	
Fig.~\ref{ROTWIRS} plots the ROTs  of the beamforming matrices of the RIS-Aided approach versus the user target SINRs with different number of BS's antennas and reflective elements of the RIS. The results confirm Theorem 1 and Corollary 1 as all ROTs are close to zero, i.e., in the order of $10^{-7}$ and $10^{-8}$, indicating  that the optimal solutions to problem \eqref{prob_IRS_W_relax} are all rank-one matrices. 
	\begin{figure}[t]
		\centering
		\includegraphics[width=.45\textwidth]{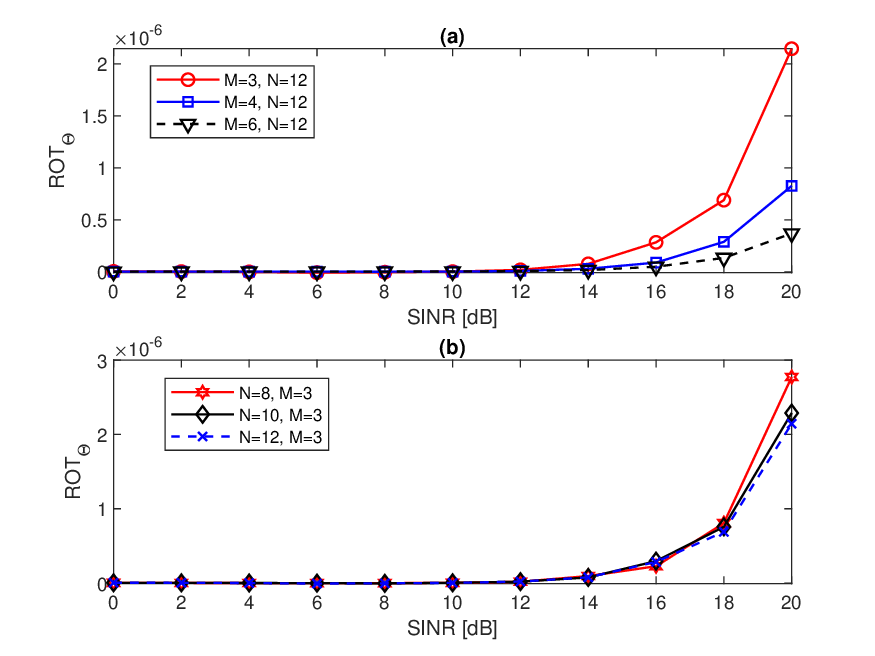} 
		\vspace*{-0.2cm}
		\caption{ROT of the phase-shift coefficient matrix versus user target SINR. (a): with different numbers of BS's antennas and a fixed number of reflective elements $N=12$. (b):  with different numbers of reflective elements and a fixed number of BS's antenna $M=3$. Uncorrelated channel model.}
		\label{ROTPh}
		\vspace*{-0.3cm}
	\end{figure}
	
Fig.~\ref{ROTPh} illustrates the ROTs of the phase-shift coefficient matrix of the RIS-Aided approach versus the user target SINRs with different number of BS's antennas and reflective elements of the RIS. Near-zero values of the ROTs shown in the figure, i.e., in the order of $10^{-6}$, imply that the optimal solutions to problems \eqref{prob_IRS_Theta_relax_std} are all rank-one matrices. This again confirms Theorem 1 and Corollary 1.  

It can be seen from  Figs.~\ref{ROTWIRS} and \ref{ROTPh} that the ROTs of the beamforming and the phase-shift coefficient matrices  are almost the same under the investigated SINR range. This implies the fact that under uncorrelated channels, the number of BS's antennas, the number of IRS's reflecting elements, and the SINR level do not have a significant impact on the rank-one property of both the beamforming and the phase-shift coefficient matrices. 

\subsubsection{Correlated Channel Model}In this experiment, Monte Carlo simulations are performed with $\numprint{500}$ independently channel realizations for each SINR target ranging from $0$ dB to $20$ dB with a step of 2. All channel realizations are feasible for the observed parameters.
\begin{figure}[t]
		\centering
		\includegraphics[width=.45\textwidth]{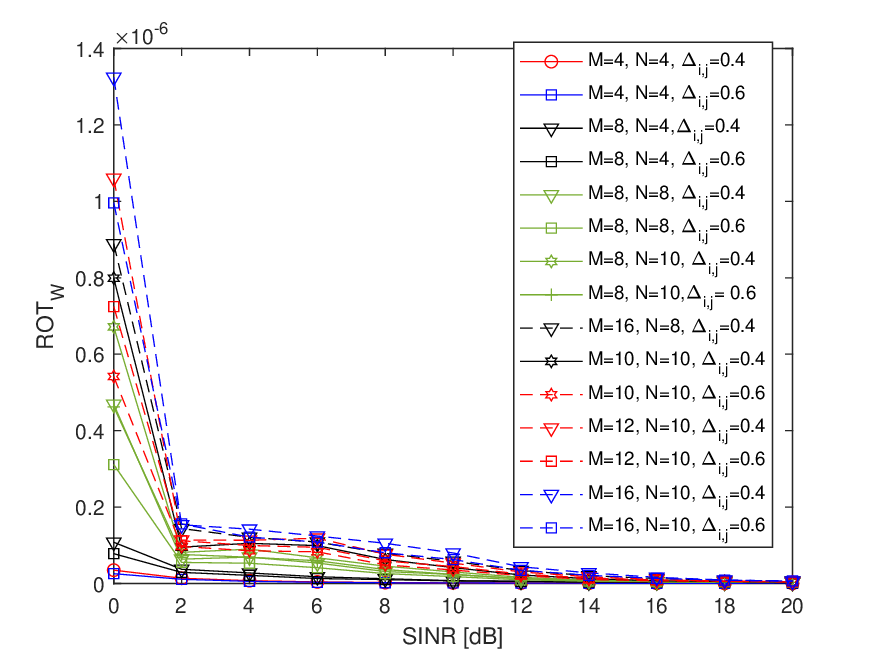} 
		\vspace*{-0.2cm}
		\caption{ROT of the beamforming matrix versus user target SINR with different numbers of BS's antennas, number of reflective elements, and values of $\boldsymbol{\Delta}_{i,j}$. Correlated channel model.}
		\label{ROTWIRS_correlated}
		\vspace*{-0.3cm}
	\end{figure}
 
Fig.~\ref{ROTWIRS_correlated}  illustrates the ROTs of the beamforming matrices of the RIS-Aided approach versus the user target SINRs with different number of BS's antennas, RIS's reflective elements and values of $\boldsymbol{\Delta}_{i,j}$. As all the ROTs are close to zeros, i.e., in the order of $10^{-6}$, Theorem 1 and Corollary 1 are confirmed. The results reveal a fact that the setup, i.e., the selection of number of BS's antennas, RIS's reflective elements and  values of $\boldsymbol{\Delta}_{i,j}$, does not make a significant impact on the ROTs of the beamforming matrices. In other words, these parameters do not have a significant impact on the ROTs of problems having $C1$ constraints.

\begin{figure}[t]
		\centering
		\includegraphics[width=.45\textwidth]{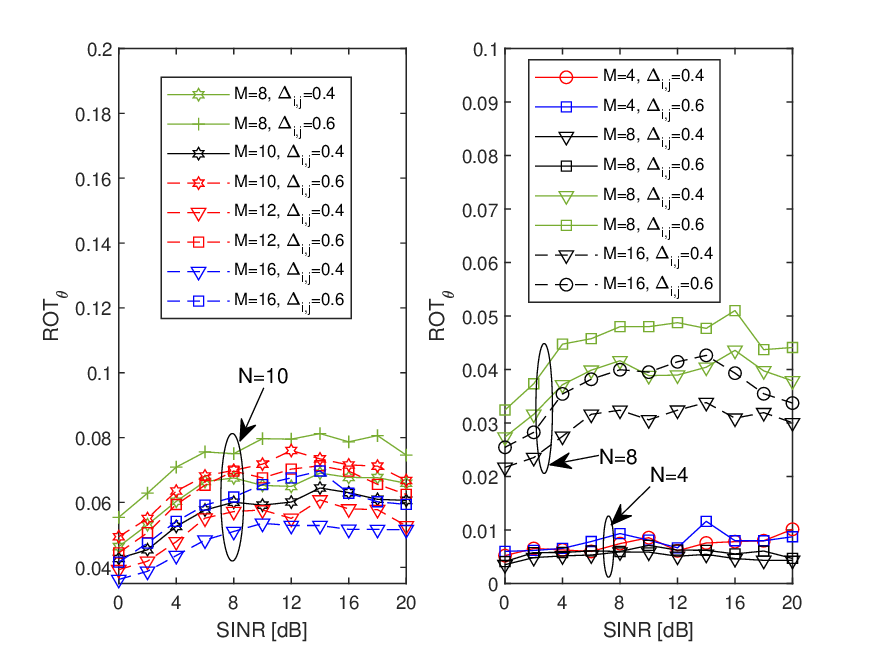} 
		\vspace*{-0.2cm}
		\caption{ROT of the phase-shift coefficient matrix versus user target SINR with different numbers of BS's antennas, number of reflective elements, and values of $\boldsymbol{\Delta}_{i,j}$. Correlated channel model.}
		\label{ROTPh_correlated}
		\vspace*{-0.3cm}
	\end{figure}

Fig.~\ref{ROTPh_correlated}  plots the ROTs of the phase-shift coefficient matrix of the RIS-Aided approach versus the user target SINRs with different number of BS's antennas, RIS's reflective elements and values of $\boldsymbol{\Delta}_{i,j}$. The results on Fig.~\ref{ROTPh_correlated} indicate that all the ROTs are close to zeros, i.e., in the order of $10^{-2}$, hence, Theorem 1 and Corollary 1 are confirmed. Fig.~\ref{ROTPh_correlated} also shows that the ROTs of the phase-shift coefficient matrix of the RIS-Aided approach increase as $\boldsymbol{\Delta}_{i,j}$ rises. This indicates a fact that optimizations problems associated with constraints $C5$ are sensitive to the values of $\boldsymbol{\Delta}_{i,j}$. The ROTs also increase as the number of reflective elements $N$ grows. For example, the ROTs are less than $0.01$ when $N=4$ while they are in the range of $0.02$ to $0.05$ when $N=8$. At the same number of reflective elements, the ROTs decrease if the number of BS's antennas increases. For instance, with $N=10$ at the SINR level of 18 dB and $\boldsymbol{\Delta}_{i,j}=0.6$, the ROTs are, respectively, 0.08, 0.071, 0.065, and 0.06 for $M=8$, $M=10$, $M=12$, and $M=16$. However, at a given number of BS's antennas, the ROTs rise as the number of reflective elements grows. For example, with $M=8$ at the SINR level of 10 dB and $\boldsymbol{\Delta}_{i,j}=0.4$, the ROTs are, respectively, 0.0058, 0.038, and 0.065 for $N=4$, $N=8$, and $N=10$.
\section{Conclusion}
We proposed an optimization framework capturing a mix of linear-matrix-inequality and second-order-cone constraints. The proposed framework can be adopted as the semidefinite relaxation counterpart of several quadratically-constrained-quadratic programs in beamforming. We analytically proved that the proposed optimization problem always yields rank-one optimal solutions if it is feasible and its optimal solution is not trivial. This in fact shows the tightness of the proposed framework when serving as a semidefinite relaxation counterpart. We took transmit beamforming as an example to demonstrate possible adoptions of the proposed framework in deriving semidefinite relaxation counterparts for quadratically-constrained-quadratic-program beamforming problems  with both perfect and imperfect channel state information. Numerical simulations have confirmed our analytical proof.
\appendices
\section{}
\begin{pro}\label{proposition1}
If an $M\times M$ Hermitian matrix $\mathbf{W}_{i}$ has a rank of $D\leq M$, then it can be expressed as
\begin{equation}
\mathbf{W}_{i}=\sum_{d=1}^{D}\lambda_{i,d}
\mathbf{w}_{i,d}\mathbf{w}_{i,d}^\text{H},\label{Propo1}
\end{equation}
where $\lambda_{i,d}$ and $\mathbf{w}_{i,d}$ are the $d$th non-zero eigenvalue and the
corresponding eigenvector of $\mathbf{W}_{i}$, respectively.
\end{pro}
\begin{proof}
As $\mathbf{W}_{i}$ has a rank of $D$, its has $D$ non-zero eigenvalues, i.e., $\lambda_{i,d}$, $d\in \{1,\cdots,D \}$, and $D$ eigenvectors, i.e., $\mathbf{w}_{i,d}$, $d\in \{1,\cdots,D \}$. Since $\mathbf{W}_{i}$ is a Hermitian matrix, its $D$  eigenvalues are real and its $D$ eigenvectors are orthogonal.  Consequently, one can express  $\mathbf{W}_{i}$ as in \eqref{Propo1}. This concludes the proof.
\end{proof}
\section{}
\begin{lemma}\label{rank_one_lemma}
Given the set of $U$ Hermitian and possitive semidefinite matrices $ \mathbf{W}_i\in \mathbb{H}^{M\times M}, \mathbf{W}_i \succeq \mathbf{0}$, $i \in\{1,2,\cdots,U\}$, and the set of $U$ possitive semidefinite matrices $\mathbf{\Phi}_{i} \succeq \mathbf{0}$, $i \in\{1,2,\cdots,U\}$, if $ \mathbf{W}_i^{\star}$ is the non-trival solution to the following problem:
\begin{eqnarray}\label{dual_func_lemma}
\displaystyle \inf_{\{\mathbf{W}_{i}\}\in \mathbb{H}^{M\times M}} \sum_{i=1}^U \textrm{Tr}\left(\mathbf{\Phi}_{i} \mathbf{W}_{i}\right),
\end{eqnarray} 
then $ \mathbf{W}_i^{\star}$ must be rank one $\forall i$.
\end{lemma}

\begin{proof} We use contradiction to prove the lemma.  We assume that the solution of \eqref{dual_func_lemma} $\mathbf{W}_{i}^{\star}$ has a rank of $D>1$, $\forall i$. Proposition \ref{proposition1} implies that
\begin{eqnarray}
\mathbf{W}_{i}^{\star}=\sum_{d=1}^D\lambda_{i,d}
\mathbf{w}_{i,d}\mathbf{w}_{i,d}^\text{H},\label{OpW}
\end{eqnarray}
where $\lambda_{i,d}\neq 0$. Since ${\mathbf{W}}_{i}^{\star}$ is also a positive semidefinite matrix, it is obvious that $\lambda_{i,d}\geq 0, \ \forall d$. Therefore, $\lambda_{i,d}> 0, \ \forall d$.

Now, we form another feasible solution to \eqref{dual_func_lemma} as
\begin{equation}\label{subop_lemma}
\widehat{\mathbf{W}}_{i}^{\star}=\lambda_{i,m} \mathbf{w}_{i,m}\mathbf{w}_{i,m}^\text{H}, \ \forall i,
\end{equation}
where
\begin{equation}
    m=\displaystyle \textrm{arg}\min_{d\in\{1,\cdots,D\}}\lambda_{i,d}
\mathbf{w}_{i,d}^\text{H}\mathbf{\Phi}_{i}\mathbf{w}_{i,d}.\label{OpWnew_lemma}
\end{equation}

As $\mathbf{\Phi}_{i}\succeq \mathbf{0}$, it is clear that 
\begin{equation}
    \lambda_{i,d}
\mathbf{w}_{i,d}^\text{H}\mathbf{\Phi}_{i}\mathbf{w}_{i,d}\geq 0, \ \forall i,d. \label{lam_lemma}
\end{equation}

Combining \eqref{OpWnew_lemma} and \eqref{lam_lemma} leads to
\begin{eqnarray}
\lambda_{i,m}
\mathbf{w}_{i,m}^\text{H}\mathbf{\Phi}_{i}\mathbf{w}_{i,m}< \sum_{d=1}^D\lambda_{i,d}
\mathbf{w}_{i,d}^\text{H}\mathbf{\Phi}_{i}\mathbf{w}_{i,d}, \ \forall i,\\
\Leftrightarrow  \textrm{Tr}\left(\mathbf{\Phi}_{i}\lambda_{i,m}\mathbf{w}_{i,m}\mathbf{w}_{i,m}^\text{H}\right)<\textrm{Tr}\left( \sum_{d=1}^D\mathbf{\Phi}_{i}\lambda_{i,d}\mathbf{w}_{i,d}\mathbf{w}_{i,d}^\text{H}\right), \ \forall i,\\
\Leftrightarrow \textrm{Tr}\left(\mathbf{\Phi}_{i} \widehat{\mathbf{W}}_{i}^{\star}\right)<
 \textrm{Tr}\left(\mathbf{\Phi}_{i} \mathbf{W}_{i}^{\star}\right), \ \forall i.
\end{eqnarray}

Therefore,
\begin{equation}\label{contradict_lemma}
\sum_{i=1}^U \textrm{Tr}\left(\mathbf{\Phi}_{i} \widehat{\mathbf{W}}_{i}^{\star}\right)<
\sum_{i=1}^U  \textrm{Tr}\left(\mathbf{\Phi}_{i} \mathbf{W}_{i}^{\star}\right).
 \end{equation}
The inequality in \eqref{contradict_lemma} contradicts the assumption that $\mathbf{W}_{i}^{\star}$ is the solution of \eqref{dual_func_lemma}. Hence, $D\leq 1$. Furthermore, assumption  $\mathbf{W}_{i}^{\star}\neq \mathbf{0}$ $\forall i$ implies that $D\geq 1$. Therefore, $D=1$. In other words, matrix $\mathbf{W}_{i}^{\star}$ must have a rank of one for all $i$.
\end{proof}
\section{Proof of Theorem \ref{theo1}}\label{apen1}
This proof is based on Lagrange duality. Particularly, as the primary problem \eqref{rootSDP} is convex, the duality gap with its dual problem is zero. Consequently, the optimal solution to the primary problem \eqref{rootSDP} can be attain from the optimal solution to the dual problem. Exploiting the result of Lemma~\ref{rank_one_lemma}, our aim is to express the dual problem of \eqref{rootSDP} in the form of \eqref{dual_func_lemma}. To that end, all the objective and constraints of \eqref{rootSDP} need to be in the forms as affine functions, i.e., LMI, of $\mathbf{W}_i$.

Our observations on the structure of \eqref{rootSDP} are as follows. The objective function, $C1$, $C2$, $C5$, and $C6$ in \eqref{rootSDP} are already affine functions of $\mathbf{W}_i$. However, $\{\mathbf{W}_i\}$ appears in $C3$ and $C4(a)$ in \eqref{rootSDP} as sub-blocks of supper matrices. In this proof, we introduce a novel technique to decompose $C3$ and $C4(a)$ in \eqref{rootSDP} into LMI of $\{\mathbf{W}_i\}$.

With $\mathbf{Y}_{i}\in \mathbb{H}^{M\times M}$, i.e., $\mathbf{Y}_{i}=\mathbf{Y}_{i}^H$, we first rewrite $C3$ in \eqref{rootSDP}
\begin{equation}
    \begin{bmatrix} \mathbf{Y}_i\mathbf{B}_i\mathbf{Y}_i+\alpha_i\mathbf{I}_M&\mathbf{Y}_i\mathbf{B}_i\mathbf{y}_i\\\ \mathbf{y}_i^H\mathbf{B}_i\mathbf{Y}_i&\mathbf{y}_i^H\mathbf{B}_i\mathbf{y}_i+
d_i
\end{bmatrix}\succeq \mathbf{0}
\end{equation}
as the following LMI constraint
\begin{equation}
\mathbf{F}(\alpha_i)+\mathbf{G}_i^H\mathbf{B}_i\mathbf{G}_i\succeq \mathbf{0},\label{LMI_1}
\end{equation}
where
\begin{eqnarray}
\mathbf{F}(\alpha_i)&=& \begin{bmatrix} \alpha_i\mathbf{I}_M&\mathbf{0}_{M \times 1}\nonumber \\ \mathbf{0}_{1 \times M}&
d_i\end{bmatrix}, \
 \mathbf{G}_i= \begin{bmatrix} \mathbf{Y}_i&\mathbf{y}_i\end{bmatrix}.
\end{eqnarray}

We continue by decomposing  the left hand side of $C4(a)$ in \eqref{rootSDP}, i.e.,
\begin{equation}
    \begin{bmatrix}\varrho_i \mathbf{I}_{M^2+M} & \begin{bmatrix}
e_i\mathbf{Z}_i\mathbf{C}_i
\mathbf{z}_i \\
\textrm{vec}\left( \mathbf{Z}_i\mathbf{C}_i\mathbf{Z}_i\right)
\end{bmatrix}\\
\begin{bmatrix}
e_i\mathbf{Z}_i\mathbf{C}_i
\mathbf{z}_i \\
\textrm{vec}\left( \mathbf{Z}_i\mathbf{C}_i\mathbf{Z}_i\right)
\end{bmatrix}^H&\varrho_i\end{bmatrix} \succeq \mathbf{0}, \label{const12new}
\end{equation}
and rewrite the constraint as
\begin{equation}\label{P22}
\mathbf{K}\left( \varrho_i\right)+\mathbf{L}_i+\mathbf{L}_i^H+\mathbf{J}_i+\mathbf{J}_i^H\succeq \mathbf{0},
\end{equation}
where
\begin{eqnarray}
\mathbf{K}\left( \varrho_i\right)&=&\begin{bmatrix}\varrho_i \mathbf{I}_{M^2+M} & {\mathbf{0}}_{(M^2+M) \times 1} \\
{\mathbf{0}}_{1 \times (M^2+M)}^H&\varrho_i\end{bmatrix},\\
\mathbf{L}_i&=&\begin{bmatrix} \mathbf{0}_{(M^2+M)\times(M^2+M)} &
{\mathbf{0}}_{(M^2+M) \times 1}\\
\begin{bmatrix}
e_i\mathbf{Z}_i\mathbf{C}_i
\mathbf{z}_i \\
{\mathbf{0}}_{M^2\times 1}
\end{bmatrix}^H&0\end{bmatrix},\\
\mathbf{J}_i&=&\begin{bmatrix} \mathbf{0}_{(M^2+M)\times (M^2+M)} & {\mathbf{0}}_{(M^2+M) \times 1}\\
\begin{bmatrix}
{\mathbf{0}}_{M\times 1}\\
\textrm{vec}\left( \mathbf{Z}_i\mathbf{C}_i\mathbf{Z}_i\right)
\end{bmatrix}^H& 0\end{bmatrix}.
\end{eqnarray}

Furthermore, matrix $\mathbf{L}_i$ can be expressed as
\begin{eqnarray}
\mathbf{L}_i&=&e_i\begin{bmatrix}
{\mathbf{0}}_{(M^2+M) \times 1}\\ 1 \end{bmatrix}\mathbf{z}_i ^H
\mathbf{C}_i \mathbf{Z}_i\begin{bmatrix} \mathbf{I}_M& {\mathbf{0}}_{M \times 1}& \cdots & {\mathbf{0}}_{M \times 1}\end{bmatrix}\nonumber \\
&=&\mathbf{T}\left( \mathbf{z}_i\right) \mathbf{C}_i \mathbf{P}\left( \mathbf{Z}_i\right),\label{theo2a3}
\end{eqnarray}
where
\begin{eqnarray}
\mathbf{T}\left( \mathbf{z}_i\right)&=&e_i\begin{bmatrix}
{\mathbf{0}}_{(M^2+M) \times 1}\\ 1 \end{bmatrix}\mathbf{z}_i ^H, \\ \mathbf{P}\left( \mathbf{Z}_i\right)& = &\mathbf{Z}_i\begin{bmatrix} \mathbf{I}_M& {\mathbf{0}}_{M\times 1}& \cdots & {\mathbf{0}}_{M \times 1}\end{bmatrix}, 
\end{eqnarray}
and $\begin{bmatrix} \mathbf{I}_M& {\mathbf{0}}_{M \times 1}& \cdots & {\mathbf{0}}_{M \times 1}\end{bmatrix}$ is an $M\times(M^2+M+1)$ matrix.

Let
$\mathbf{U}_i=\begin{bmatrix}
{\mathbf{0}}_{M\times 1}&\cdots &{\mathbf{0}}_{M\times 1}& \mathbf{u}_i \end{bmatrix}^T$ present an $(M^2+M+1)\times M$ matrix in which an $M \times 1$ vector $\mathbf{u}_i$ contains all zeros but $1$ at the $i$-th entry. Moreover, let $\mathbf{V}_i=\begin{bmatrix}
\underbrace{\mathbf{0}_{M \times M}}_{1\textrm{st}}&\mathbf{0}_{M \times M}&\cdots&\underbrace{\mathbf{I}_M}_{i\textrm{th}}&\cdots& \underbrace{\mathbf{0}_{M\times M}}_{(M^2+M)\textrm{th}}& {\mathbf{0}}_{M \times 1} \end{bmatrix}$ present an $M\times(M^2+M+1)$ matrix with all zeros but $\mathbf{I}_M$ at the $i$-th block. We can rewrite $\mathbf{J}_i$ as
\begin{eqnarray}
\mathbf{J}_i=\sum_{p=1}^M \mathbf{U}_p \mathbf{Z}_i\mathbf{C}_i\mathbf{Z}_i\mathbf{V}_p.\label{theo2a4}
\end{eqnarray}

Using  \eqref{P22}, \eqref{theo2a3}, and \eqref{theo2a4}, one can equivalently cast constraint \eqref{const12new} as the following LMI constraint
\begin{eqnarray}\label{const12C}
&{}&\mathbf{K}\left( \varrho_i\right)+\mathbf{T}\left( \mathbf{z}_i\right) \mathbf{C}_i \mathbf{P}\left( \mathbf{Z}_i\right)+\left[\mathbf{T}\left( \mathbf{z}_i\right) \mathbf{C}_i \mathbf{P}\left( \mathbf{Z}_i\right)\right]^H
\nonumber \\ &{}&+\sum_{p=1}^M \mathbf{U}_p \mathbf{Z}_i\mathbf{C}_i\mathbf{Z}_i\mathbf{V}_p+\left[\sum_{p=1}^M \mathbf{U}_p \mathbf{Z}_i\mathbf{C}_i\mathbf{Z}_i\mathbf{V}_p\right]^H \succeq \mathbf{0}.
\end{eqnarray}

Using \eqref{LMI_1} and \eqref{const12C}, we recast \eqref{rootSDP} as:
\small
\begin{equation}
\begin{aligned}\label{transformedSDP}
& \displaystyle \min_{\{\mathbf{W}_{i}\}\in \mathbb{H}^{M\times M}, \alpha_i\geq 0, \varrho_i, f_i\geq 0} & &
\sum_{i=1}^U \textrm{Tr}\left(\mathbf{A}_{i}\mathbf{W}_{i}\right) \\
& \text{s.\ t.}\ & &a_i \textrm{Tr}\left(\mathbf{X}_{i,i}\mathbf{W}_i\right) +\sum_{j=1,j \neq i}^U b_j\textrm{Tr}\left(\mathbf{X}_{i,j}\mathbf{W}_j\right)+c_i\geq 0, \\ 
&&& \ \ \ \ \ \ \ \ \ \ \ \ \ \ \ \ \ \ \ \ \ \ \ \ \ \ \  \ \forall i \in \{1,\cdots,L_1\},\\
&&&m_i\textrm{Tr}\left(\mathbf{M}_{i}\mathbf{W}_i\right) +p_i\geq 0,\\
&&& \ \ \ \ \ \ \ \ \ \ \ \ \ \ \ \ \ \ \ \ \ \ \ \ \ \ \  \ \forall i \in \{1,\cdots,U\},\\
&&& \mathbf{F}(\alpha_i)+\mathbf{G}_i^H\mathbf{B}_i\mathbf{G}_i\succeq \mathbf{0}, \ \forall i  \in \{1,\cdots,L_3\},\\
&&& \mathbf{K}\left( \varrho_i\right)+\mathbf{T}\left( \mathbf{z}_i\right) \mathbf{C}_i \mathbf{P}\left( \mathbf{Z}_i\right)+\left[\mathbf{T}\left( \mathbf{z}_i\right) \mathbf{C}_i \mathbf{P}\left( \mathbf{Z}_i\right)\right]^H\\
&&&+\sum_{p=1}^M \mathbf{U}_p \mathbf{Z}_i\mathbf{C}_i\mathbf{Z}_i\mathbf{V}_p+\left[\sum_{p=1}^M \mathbf{U}_p \mathbf{Z}_i\mathbf{C}_i\mathbf{Z}_i\mathbf{V}_p\right]^H \succeq \mathbf{0},\\ 
&&& \ \ \ \ \ \ \ \ \ \ \ \ \ \ \ \ \ \ \ \ \ \ \ \ \ \ \   \forall i  \in \{1,\cdots,L_4\},\\
&&& f_i\mathbf{I}_N+v_i\mathbf{D}_i\sum_{k=1}^N\boldsymbol\Lambda_k\mathbf{E}_i\boldsymbol\Psi_k\widetilde{\mathbf{D}}_i\succeq \mathbf{0},\ \forall i  \in \{1,\cdots,L_5\},\\
&&& \mathbf{W}_{i} \succeq \mathbf{0}, \ \forall i \in \{1,\cdots,U\}.
\end{aligned}
\end{equation}
\normalsize
\normalsize
It is clear that problem \eqref{transformedSDP} is convex as it is in a SDP form \cite{Lieven}. Since problem \eqref{transformedSDP} is convex and satisfies Slater's constraint qualification \cite{Boyd_convex}, strong duality holds. Consequently, the optimal solution of the primary problem \eqref{transformedSDP} can be attained via solving its dual problem. 
In the following, we exploit the duality to investigate the property of the optimal beamforming matrix. We proceed by establishing the Lagrangian of \eqref{transformedSDP} as
\begin{eqnarray}
\mathfrak{L}\left(\{\mathbf{W}_{i}\},\boldsymbol\Upsilon \right)&=&\sum_{i=1}^U\textrm{Tr}\left(\mathbf{A}_{i}\mathbf{W}_{i}\right)\nonumber \\&{}&-
\sum_{i=1}^{L_1} \beta_i \left(a_i \textrm{Tr}\left(\mathbf{X}_{i,i}\mathbf{W}_i\right) +\sum_{j=1}^U b_j\textrm{Tr}\left(\mathbf{X}_{i,j}\mathbf{W}_j\right)+c_i\right)\nonumber\\
&{}&-\sum_{i=1}^{U}\tau_i\left( m_i\textrm{Tr}\left(\mathbf{M}_{i}\mathbf{W}_i\right) +p_i\right)\nonumber\\
&{}&-\sum_{i=1}^{L_3}\textrm{Tr}\left(
\mathbf{Q}_i\left[ \mathbf{F}(\alpha_i)+\mathbf{G}_i^H\mathbf{B}_i\mathbf{G}_i\right]\right)
-\sum_{i=1}^{L_3}\kappa_i \alpha_i\nonumber \\
&{}&-\sum_{i=1}^{L_4}\textrm{Tr}\left(\mathbf{R}_i \mathbf{\Xi}_{i} \right)\nonumber\\
&{}&-\sum_{i=1}^{L_5}\textrm{Tr}\left( \mathbf{S}_i\left[f_i\mathbf{I}_N+v_i\mathbf{D}_i\sum_{k=1}^N\boldsymbol\Lambda_k\mathbf{E}_i\boldsymbol\Psi_k\widetilde{\mathbf{D}}_i \right]\right) \nonumber \\
&{}&-\sum_{i=1}^U\textrm{Tr}\left( \mathbf{N}_i \mathbf{W}_i\right)
,\label{lagran}
\end{eqnarray}
\normalsize
where 
\begin{eqnarray}
\mathbf{\Xi}_{i}&=&\mathbf{K}\left( \varrho_i\right)+\mathbf{T}\left( \mathbf{z}_i\right) \mathbf{C}_i \mathbf{P}\left( \mathbf{Z}_i\right)+\left[\mathbf{T}\left( \mathbf{z}_i\right) \mathbf{C}_i \mathbf{P}\left( \mathbf{Z}_i\right)\right]^H\nonumber \\
&{}&+\sum_{p=1}^M \mathbf{U}_p \mathbf{Z}_i\mathbf{C}_i\mathbf{Z}_i\mathbf{V}_p+\left[\sum_{p=1}^M \mathbf{U}_p \mathbf{Z}_i\mathbf{C}_i\mathbf{Z}_i\mathbf{V}_p\right]^H,
\end{eqnarray}
$\beta_i\geq 0$, $\tau_i\geq 0$, $\mathbf{Q}_i\succeq \mathbf{0}$, $\kappa_i\geq 0$, $\mathbf{R}_i\succeq \mathbf{0}$, $\mathbf{S}_i\succeq \mathbf{0}$, and $\mathbf{N}_i\succeq \mathbf{0}$ are the Lagrange multipliers associated with the constraints in \eqref{transformedSDP}, respectively. We represent these Lagrange multipliers in compact forms as follows: $\boldsymbol\alpha=\begin{bmatrix} \alpha_{1},\cdots,\alpha_{L_3}\end{bmatrix}^T$,
$\boldsymbol\beta=\begin{bmatrix} \beta_{1},\cdots,\beta_{L_1}\end{bmatrix}^T$,
$\boldsymbol\kappa=\begin{bmatrix} \kappa_{1},\cdots,\kappa_{L_3}\end{bmatrix}^T$, $\boldsymbol\tau=\begin{bmatrix}\tau_{1},\cdots,\tau_{U} \end{bmatrix}^T$,  $\{\mathbf{Q}_{i}\}=\{\mathbf{Q}_{1},\cdots,\mathbf{Q}_{L_3} \}$,
$\{\mathbf{R}_{i}\}=\{\mathbf{R}_{1},\cdots,\mathbf{R}_{L_4}\}$,
$\{\mathbf{S}_i\}=\{\mathbf{S}_{1},\cdots,\mathbf{S}_{L_5}\}$,
$\{\mathbf{N}_{i}\}=\{\mathbf{N}_{1},\cdots,\mathbf{N}_{U} \}$, and  $\boldsymbol\Upsilon=\left\{
\boldsymbol\alpha,\boldsymbol\beta,\boldsymbol\tau,\boldsymbol\kappa,\{\mathbf{Q}_{i}\},\{\mathbf{R}_{i}\},
\{\mathbf{S}_{i}\},\{\mathbf{N}_{i}\}\right\}
$ as the set of the dual variables.
 
Consider the dual function of \eqref{transformedSDP} as 
\begin{eqnarray}\label{dual_func1}
g\left(\boldsymbol\Upsilon \right)=\displaystyle \inf_{\{\mathbf{W}_{i}\}\in \mathbb{H}^{M\times M}}
\mathfrak{L}\left(\{\mathbf{W}_{i}\},\boldsymbol\Upsilon\right). \end{eqnarray}
The corresponding dual problem of \eqref{transformedSDP} is then expressed as
\begin{equation}\label{dual}
\begin{aligned}
&\displaystyle \max_{\boldsymbol\Upsilon}
 &&g\left(\boldsymbol\Upsilon \right)\\
& \text{s.\ t.}\ & &\boldsymbol\alpha\succcurlyeq \mathbf{0},\boldsymbol\beta \succcurlyeq \mathbf{0}, \boldsymbol\kappa\succcurlyeq \mathbf{0}, \\&&& \mathbf{Q}_{i} \succeq \mathbf{0},\mathbf{R}_i \succeq \mathbf{0},\  \mathbf{S}_{i} \succeq \mathbf{0},\mathbf{N}_i \succeq \mathbf{0},\ \forall i.
\end{aligned}
\end{equation}

In the sequel, the optimal solution of the primary problem \eqref{transformedSDP} is attained via solving its dual problem \eqref{dual}. To that end, let $\boldsymbol\Upsilon^{\star}=\{
\boldsymbol\alpha^{\star},\boldsymbol\beta^{\star},\boldsymbol\tau^{\star},\boldsymbol\kappa^{\star},\{\mathbf{Q}_{i}^{\star}\},\{\mathbf{R}_{i}^{\star}\},\{\mathbf{S}_{i}^{\star}\},
\{\mathbf{N}_{i}^{\star}\}\}$ represent the optimal solution to the dual problem \eqref{dual}, then the corresponding
optimal solution $\{\mathbf{W}_{i}^{\star}\}$ to the primary problem \eqref{transformedSDP}
can be attained as
\begin{eqnarray}\label{dual_func}
g(\boldsymbol\Upsilon^{\star})=\displaystyle \inf_{\{\mathbf{W}_{i}\}\in \mathbb{H}^{M\times M}}
\mathfrak{L}\left(\{\mathbf{W}_{i}\},\boldsymbol\Upsilon^{\star}\right).
\end{eqnarray}

Substituting for $\mathbf{B}_i=g_i\mathbf{W}_i+\sum_{j=1}^U h_j\mathbf{W}_j$, $\mathbf{C}_i=\tilde{g}_i\mathbf{W}_i+\sum_{j=1}^U \tilde{h}_j\mathbf{W}_j$, $\mathbf{E}_i=\bar{g}_i\mathbf{W}_i+\sum_{j=1}^U \bar{h}_j\mathbf{W}_j$ in \eqref{lagran} with $a_i=0$, $g_i=0$, $\tilde{g}_i=0$, $\bar{g}_i=0$,  $\forall i> U$, after some mathematical manipulations, we can express \eqref{lagran} as
 \begin{eqnarray}
 \mathfrak{L}\left(\{\mathbf{W}_{i}\},\boldsymbol\Upsilon^{\star} \right)= \sum_{i=1}^U \textrm{Tr}\left(\mathbf{\Phi}_{i} \mathbf{W}_{i}\right)+\eta,
 \end{eqnarray}
 where 
\begin{eqnarray}
\mathbf{\Phi}_{i}&=&\mathbf{A}_i-\beta_i^{\star} a_i \mathbf{X}_{i,i}-b_i\sum_{j=1}^{L_1}\mathbf{X}_j-\tau_{i}^{\star}m_i\mathbf{M}_i\nonumber \\&{}&-g_i\mathbf{G}_i\mathbf{Q}_i^{\star}\mathbf{G}_i^H-h_i\sum_{j=1}^{L_3}\mathbf{G}_j\mathbf{Q}_j^{\star}\mathbf{G}_j^H \nonumber \\
&{}&-\tilde{g}_i\mathbf{P}(\mathbf{Z}_i)\mathbf{R}_i^{\star}\mathbf{T}(\mathbf{z}_i)-\tilde{h}_i\sum_{j=1}^{L_4}\mathbf{P}(\mathbf{Z}_j)\mathbf{R}_j^{\star}\mathbf{T}(\mathbf{z}_j)\nonumber \\
&{}&-\tilde{g}_i\mathbf{T}^H(\mathbf{z}_i)\mathbf{R}_i^{\star}\mathbf{P}^H(\mathbf{Z}_i)-\tilde{h}_i\sum_{j=1}^{L_4}\mathbf{T}^H(\mathbf{z}_j)\mathbf{R}_j^{\star}\mathbf{P}^H(\mathbf{Z}_j)\nonumber \\
&{}&-\tilde{g}_i\sum_{p=1}^{M}\mathbf{Z}_i\mathbf{V}_p\mathbf{R}_i^{\star}\mathbf{U}_p\mathbf{Z}_i-\tilde{h}_i\sum_{j=1}^{L_4}\sum_{p=1}^{M}\mathbf{Z}_j\mathbf{V}_p\mathbf{R}_j^{\star}\mathbf{U}_p\mathbf{Z}_j\nonumber \\
&{}&-\tilde{g}_i\sum_{p=1}^{M}\mathbf{Z}_i^H\mathbf{U}^H_p\mathbf{R}_i^{\star}\mathbf{V}^H_p\mathbf{Z}_i^H-\tilde{h}_i\sum_{j=1}^{L_4}\sum_{p=1}^{M}\mathbf{Z}^H_j\mathbf{U}^H_p\mathbf{R}_j^{\star}\mathbf{V}^H_p\mathbf{Z}^H_j\nonumber \\
&{}&-\bar{g}_i v_i\sum_{k=1}^{N}\boldsymbol\Psi_k\widetilde{\mathbf{D}}_i\mathbf{S}_i^{\star}\mathbf{D}_i\boldsymbol\Lambda_k-\bar{h}_i v_i \sum_{j=1}^{L_5}\sum_{k=1}^{N}\boldsymbol\Psi_k\widetilde{\mathbf{D}}_j\mathbf{S}_j^{\star}\mathbf{D}_j\boldsymbol\Lambda_k-\mathbf{N}_i,\label{phi_con}\nonumber \\
\end{eqnarray}
and 
\begin{eqnarray}
\eta&=&-\sum_{i=1}^{L_1}\beta_{i}^{\star}c_i-\sum_{i=1}^{U}\tau_{i}^{\star}p_i-\sum_{i=1}^{L_3}\textrm{Tr}\left(\mathbf{Q}_i^{\star}\mathbf{F}\left( \alpha_i^{\star}\right) \right)-\sum_{i=1}^{L_3}\kappa_i^{\star}\alpha_i^{\star}\nonumber \\
&{}&-\sum_{i=1}^{L_4}\textrm{Tr}\left(\mathbf{R}_i^{\star}\mathbf{K}\left( \varrho_i\right) \right)-\sum_{i=1}^{L_5}\textrm{Tr}\left(f_i^{\star}\mathbf{S}_i^{\star}\right).
\end{eqnarray}

At the optimal point, it is clear that $\eta$ is a constant. Hence, we can equivalently cast \eqref{dual_func} as
\begin{eqnarray}\label{dual_func2}
g(\boldsymbol\Upsilon^{\star})
=\displaystyle \inf_{\{\mathbf{W}_{i}\}\in \mathbb{H}^{M\times M}} \sum_{i=1}^U \textrm{Tr}\left(\mathbf{\Phi}_{i} \mathbf{W}_{i}\right).
\end{eqnarray}

We have decomposed two SOC constraints $C3$ and $C4(a)$ in \eqref{rootSDP}, respectively, into two LMIs of $\mathbf{W}_i$ as \eqref{LMI_1} and \eqref{const12C}. Unlike the methods in \cite{zheng2016} and \cite{KhandakerSep2016}, our novel SOC decomposition method does not
reduce the feasibility region of the transformed SDRs. We have then equivalently expressed \eqref{rootSDP} as \eqref{transformedSDP}.  The technique used to transform \eqref{rootSDP} into \eqref{transformedSDP} is the generalization of our previous works in \cite{TuanWCL2015,Tuanglobecom15no1}, and \cite{TuanTGCN2017}.  If \eqref{rootSDP} is feasible, then its equivalent form \eqref{transformedSDP} is also feasible. Therefore, the optimal value of  \eqref{transformedSDP} is non-negative.\footnote{Since \eqref{rootSDP} and its equivalent form \eqref{transformedSDP} are both convex and feasible, a unique optimal solution exits within the feasible region and can be efficiently obtained by interior-point methods, e.g., using CVX package.} Moreover, the duality gap between the primary problem \eqref{transformedSDP} and its Lagrange dual problem \eqref{dual} is zero. Consequently, matrix $\mathbf{\Phi}_{i}$  must be positive semi-definite, i.e.,
$\mathbf{\Phi}_{i}\succeq \mathbf{0}, \ \forall i$,  such that the Lagrangian dual function is bounded from below.\footnote{The conditions for the assumption that \eqref{rootSDP} is feasible, i.e., $\mathbf{\Phi}_{i}\succeq \mathbf{0}, \ \forall i$, are out of the scope of this paper. Such conditions may relate to the structure of the primary problem \eqref{rootSDP}, i.e., the parameters of the matrix $\mathbf{\Phi}_{i}$ in \eqref{phi_con}, for examples, the relationship between the number of constraints and the number of variables, or the input data of the problem, or the severeness of channel estimation errors. This investigation deserves further research.}

Let $\{\mathbf{W}_i^{\star}\}=\{\mathbf{W}_1^{\star},\mathbf{W}_2^{\star},\cdots,\mathbf{W}_U^{\star}\}$ be the non trivial optimal solution of the primary problem \eqref{rootSDP}. Consequently, $\{\mathbf{W}_{i}^{\star}\}$ is also the solution of \eqref{dual_func2}. Since $\mathbf{\Phi}_{i}\succeq \mathbf{0}, \ \forall i$, Lemma~\eqref{rank_one_lemma} indicates that matrix $\mathbf{W}_{i}^{\star}$ must have a rank of one for all $i$. This concludes the proof. $\blacksquare$ 
\bibliographystyle{IEEEtran}
\bibliography{VT_2022_04576_Camera_Ready}
\begin{IEEEbiography}[{\includegraphics[width=1in,height=1.25in]{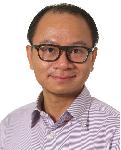}}]{Tuan Anh Le (S'10-M'13-SM'19)} received the Ph.D. degree in telecommunications research from
King’s College London, The University of London, U.K., in 2012. He was a Post-Doctoral Research Fellow with the School of Electronic and
Electrical Engineering, University of Leeds, Leeds, U.K. He is a Senior Lecturer at Middlesex University, London, U.K. His current research interests include integrated sensing and communication (ISAC), RIS-aided communication, RF energy harvesting and wireless power transfer, physical-layer security, and applied machine learning for wireless communications. He severed as a Technical Program Chair for 26th International Conference on Telecommunications (ICT 2019). He was an Exemplary Reviewer of IEEE Communications Letters in 2019.
\end{IEEEbiography}
\begin{IEEEbiography}[{\includegraphics[width=1in,height=1.25in]{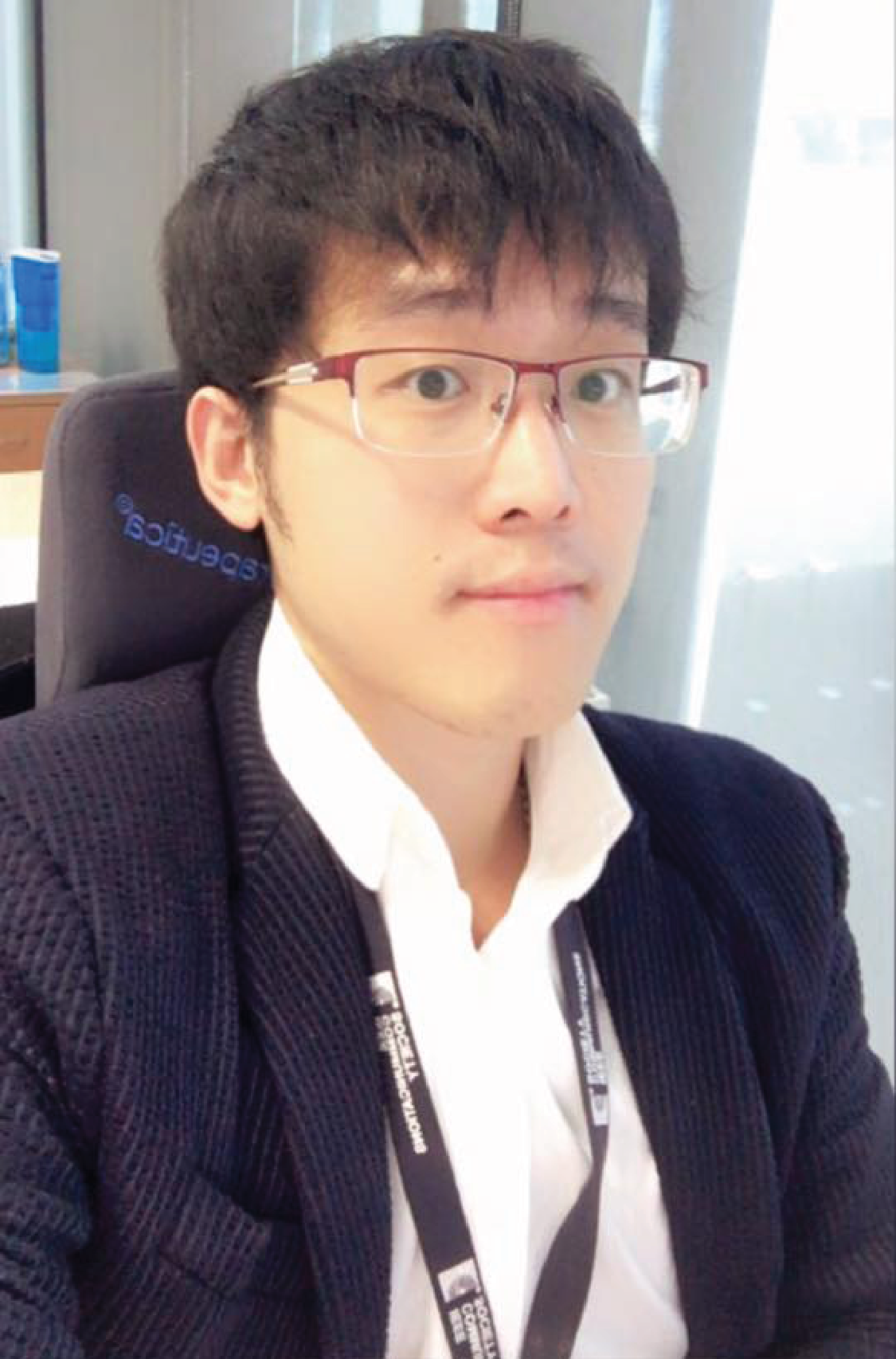}}]{Derrick Wing Kwan Ng (S'06-M'12-SM'17-F'21)} received a bachelor's degree with first-class honors and a Master of Philosophy (M.Phil.) degree in electronic engineering from the Hong Kong University of Science and Technology (HKUST) in 2006 and 2008, respectively. He received his Ph.D. degree from the University of British Columbia (UBC) in Nov. 2012. He was a senior postdoctoral fellow at the Institute for Digital Communications, Friedrich-Alexander-University Erlangen-N\"urnberg (FAU), Germany. He is now working as a Scientia Associate Professor at the University of New South Wales, Sydney, Australia. His research interests include global optimization, physical layer security, IRS-assisted communication, UAV-assisted communication, wireless information and power transfer, and green (energy-efficient) wireless communications.

Dr. Ng has been listed as a Highly Cited Researcher by Clarivate Analytics (Web of Science) since 2018. He received the Australian Research Council (ARC) Discovery Early Career Researcher Award 2017, the IEEE Communications Society Leonard G. Abraham Prize 2023, the IEEE Communications Society Stephen O. Rice Prize 2022, the Best Paper Awards at the WCSP 2020, 2021, IEEE TCGCC Best Journal Paper Award 2018, INISCOM 2018, IEEE International Conference on Communications (ICC) 2018, 2021, 2023, IEEE International Conference on Computing, Networking and Communications (ICNC) 2016, IEEE Wireless Communications and Networking Conference (WCNC) 2012, the IEEE Global Telecommunication Conference (Globecom) 2011, 2021 and the IEEE Third International Conference on Communications and Networking in China 2008. He served as an editorial assistant to the Editor-in-Chief of the IEEE Transactions on Communications from Jan. 2012 to Dec. 2019. He is now serving as an editor for the IEEE Transactions on Communications and an Associate Editor-in-Chief for the IEEE Open Journal of the Communications Society.
\end{IEEEbiography}
\begin{IEEEbiography}[{\includegraphics[width=1in,height=1.25in]{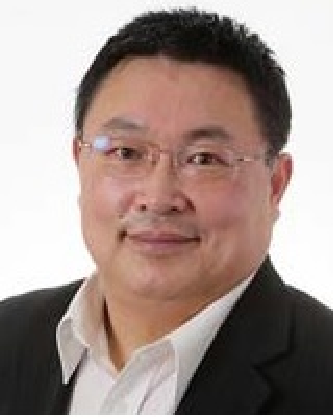}}]{Xin-She Yang} obtained his DPhil in Applied Mathematics from the
University of Oxford. He then worked at Cambridge University and
National Physical Laboratory (UK) as a Senior Research Scientist.
Now he is Reader at Middlesex University London, and a co-Editor
of the Springer Tracts in Nature-Inspired Computing. He is also
an elected Fellow of the Institute of Mathematics and its
Applications. He was the IEEE Computational Intelligence
Society (CIS) chair for the Task Force on Business Intelligence
and Knowledge Management (2015 to 2020). He has published
more than 300 peer-reviewed research papers with more
than 82,000 citations, and he has been on the prestigious
list of highly-cited researchers (Web of Sciences) for
seven consecutive years (2016-2022).
\end{IEEEbiography}
\end{document}